\documentclass[preprint]{aastex}
\usepackage[plainpages=false,  pdfpagelabels,  naturalnames=true,pdftex,linktocpage]{hyperref}
\hypersetup{
  pdfpagemode=None, 		
  pdfstartview=Fit,				
  pdfproducer = {pdflatex}}
\usepackage{amsmath} 
\usepackage{sistyle}		
\usepackage[dvips]{epsfig}
\usepackage[dvips]{graphicx}
\usepackage[dvips]{color}						
\begin{document}
%%%%%%%%%%%%%%%%%%%%%%%%%%%%%%%%%%%%%%%%%%%%%%%%
%%%%%%%%%%%%%%%%%%%%%%%%%%%%%%%%%%%%%%%%%%%%%%%%
\title{Internal $\gamma\gamma$-opacity in Active Galactic Nuclei and the consequences for the TeV observations of M87 and Cen$\,$A}
\shorttitle{Internal $\gamma\gamma$-opacity in M87 and Cen\,A }
\author{Katharina A.~Brodatzki\altaffilmark{1}\altaffilmark{*}, David J.~S.~Pardy\altaffilmark{2}, Julia K.~Becker\altaffilmark{1},
   and Reinhard Schlickeiser\altaffilmark{1}}
\shortauthors{K.~A.~Brodatzki}
\altaffiltext{1}{Institut f\"ur Theoretische Physik, Lehrstuhl IV: Weltraum- und Astrophysik, Ruhr-Universität Bochum, D-44780 Bochum, Germany}
\altaffiltext{2}{Queen's University, Kingston, Ontario, K7L 3N6, Canada}
\altaffiltext{*}{Contact: kb@tp4.rub.de}
%%%%%%%%%%%%%%%%%%%%%%%%%%%%%%%%%%%%%%%%%%%%%%%%%  
%%%%%%%%%%%%%%%%%%%%%%%%%%%%%%%%%%%%%%%%%%%%%%%%%  
\begin{abstract}
{\it Low Luminosity Active Galactic Nuclei (LLAGNs) possess the characteristic features of more luminous Active Galactic Nuclei (AGNs) but exhibit a much lower nuclear H$\alpha$ luminosity \citep[$L_{H \alpha} <10^{40} \text{erg s}^{-1}$,][]{HFS97} than their more luminous counterparts. M87 (NGC 4486) and Centaurus A (NGC 5128, Cen\,A) are well-studied nearby LLAGNs. As an additional feature they show $\gamma$-radiation up to TeV ($10^{12}$eV) energies, but the origin of this radiation is not resolved. The coincident observation of a radio and TeV flare in M87 suggests that the TeV radiation is produced within around $50-100$ gravitational radii of the central supermassive black hole, depending on the assumed value of the mass of the black hole. Strong radiation fields can be produced in the central region of an (LL)AGN, e.g., by the accretion flow around the black hole, the jet plasma, or stars closely orbiting the black hole. These radiation fields can lead to the absorption of emitted TeV photons, and in fact high optical depths of such fields can make TeV detection from inner regions impossible. In this paper we consider the accretion flow around the black hole as the most prominent source for such a radiation field and we accordingly calculate the probability for absorption of TeV photons produced near the black holes in M87 and CenA assuming a low luminosity Shakura-Sunyaev Disk (SSD). We find that the results are very different for between the two LLAGNs. While the inner region of M87 is transparent for TeV radiation up to $\sim 15$TeV, the optical depth in Cen\,A is $ \gg 1$, leading to an absorption of TeV photons that might be produced near the central black hole. These results imply either that the TeV $\gamma$ production sites and processes are different for both sources, or that LLAGN black holes do not accrete (at least only) in form of a low luminosity SSD.}
\end{abstract}
\parindent=0cm
\parskip=0.2cm
%%%%%%%%%%%%%%%%%%%%%%%%%%%%%%%%%%%%%%%%%%%%%%%%%  
%%%%%%%%%%%%%%%%%%%%%%%%%%%%%%%%%%%%%%%%%%%%%%%%%
\section{Introduction}
%%%%%%%%%%%%%%%%%%%%%%%%%%%%%%%%%%%%%%%%%%%%%%%%%
The current most favored model for Active Galactic Nuclei (AGNs), which are highly active, very compact regions in galactic centers, is the 'Unified AGN model' developed by \cite{UrryPadovani95}. In this model, many differences in the spectra of AGNs arise as a consequence of their orientation in Earth's view rather than from different physical properties. The most characteristic component of an AGN is the central black hole of the host galaxy. Strong gravitational tides attract surrounding material to form the accretion flow around the black hole, and a dust torus is formed surrounding the accretion flow. Additionally, radio-loud galaxies typically have relativistic jets which transport plasma particles from black hole vicinity to outer regions of the galaxy. Depending on the orientation of an AGN towards the Earth, some regions of the galaxy are observable whereas others are obscured. Consequently, the spectral energy distributions (SEDs) of some AGNs show features such as broad or narrow emission lines or a highly luminous accretion disk, whereas others do not.

For a long time blazars were the only known extragalactic sources that seemed to emit $\gamma$-radiation up to TeV ($10^{12}$eV) energies. The angle between the line of sight and the relativistic jet of blazars is very small, meaning the jet points towards Earth. Resultantly, radiation emitted by relativistic particles along the jet is strongly influenced by relativistic effects; on Earth we observe the emitted photons at higher energies and intensities than they actually have inside the jet. This makes blazars detectable at increasingly high energies. 

The detection of TeV $\gamma$-radiation from the radio galaxies M87 \citep[NGC 4486, ][]{Aharonian03, Aharonian06} and Centaurus A \citep[NGC 5128 and henceforth Cen\,A, ][]{Aharonian09} revealed another class of extragalactic TeV $\gamma$-ray sources. These two objects are also AGNs but their jets make larger angles between with our line of sight. For such non-blazar AGNs the observed TeV energies of the photons cannot arise from relativistic effects but must be produced inside the source. However, both their production mechanism and site are still unknown. Nevertheless, recent observations of M87 hint that the TeV photons are produced near the black hole \citep{Acciari09}. 

M87 and Cen$\,$A are both Fanaroff Riley I \citep[FR I, ][]{FanaroffRiley74} radio galaxies, meaning radio-loud but optically underluminous \citep[$M_V > -23$, ][]{Melia2009} AGNs in centers of (almost exclusively) elliptical galaxies. Due to their low nuclear luminosity they belong to the class of Low Luminosity Active Galactic Nuclei (LLAGNs).

LLAGNs have similar characteristic features to luminous AGNs but exhibit nuclear luminosities which are orders of magnitude lower than their more luminous counterparts. Interestingly, LLAGNs are oftentimes radio-loud and in several cases host relativistic radio-jets \citep{Ho2008, Nagar02}. The existence of jets in LLAGNs could indicate a connection between their low luminosity and the formation of outflows. Furthermore, the ``big blue bump'', meaning the optical-to-UV radiation which is presumably the emission from a geometrically thin yet optically thick disk common to luminous AGNs, seems to be absent in some LLAGNs \citep{Ho2008}. The lack of this thin disk characteristic favors a different accretion model than the thin disk in luminous AGNs. Due to the low luminosity nature of such objects the accretion could be explained by the Radiatively Inefficient Accretion Flow (RIAF) model in which the energy of accreted particles is either advected with the matter into the black hole \citep[the so called Advection Dominated Accretion Flow, ADAF,][]{NarayanYi94, NarayanYi95a, NarayanYi95b} or redirected into an outflow instead of being radiated away. Accretion flows which form outflows assort well with the radio loudness and the occurrence of jets in LLAGNs. Such RIAFs have rather a geometrical form of a sphere and hence a much lower optical depth than geometrically thin disks. 

On the other hand, there is also evidence that at least part of the accretion flow in a typical LLAGN is in a thin disk \citep{Maoz07}. In this paper we scrutinize the accretion form in LLAGNs with the aid of observed TeV $\gamma$-radiation from M87 and Cen\,A. Due to the different physical properties in the sundry accretion flows (like proton or electron density, the temperature of the particles, the degree of interaction between them, the type of radiation production etc.), the interaction of the accretion flow with TeV photons produced in its vicinity vary widely between different flows. The comparison of absorption probability with observations could point out which type of accretion flow can be ruled out for LLAGNs. In this paper we assume that the accretion in M87 and Cen\,A is in the form of a thin disk and accordingly calculate the strength of TeV photon absorption, with production sites at varying distances above the black hole, by disk photons. M87 and Cen\,A are excellent objects for our investigation since the mass of their black holes and the observed bolometric luminosities are quite different. These variables hold much weight in the formation of accretion flows and therefore greatly affect the absorption probability of TeV photons. Models which are very sensitive to black hole mass and bolometric luminosity could have problems forecasting the absorption probability for both  M87 and Cen\,A, since their relevant parameters vary considerably.

\subsection*{Notation used in this paper}
Throughout the paper we use the following notation:
\begin{itemize}
\item $m=M/M_{\odot}$ is the black hole mass in units of the solar mass $M_{\odot}$
\item $r=R/R_G$ is the radius of a ring segment of an accretion disk around a black hole in units of gravitational radius for the black hole 
\item $\dot{m}_{Edd}=\dot{M}/\dot{M}_{Edd}$ is the mass accretion rate in units of the Eddington accretion rate 
\item $\epsilon = E/m_e\,c^2$ is the energy of a photon in units of the rest energy of an electron, where $m_e$ denotes the electron rest mass and $c$ is the speed of light in vacuum
\end{itemize}

%-----------------------------------------------------------------------------------------------------------------------------
\begin{table}[]
\small
    \begin{tabular}{p{4.4cm}p{5.5cm}p{5.cm}}
    \hline \hline
   		  					 & 						M87 &						 Cen\,A \\ \hline
  	 & & \\
	black hole mass $m=M/M_\odot$  & $(3.2 \pm 0.9)\cdot {10^9}^{(1)}$, $(6.4 \pm 0.5)\cdot {10^9}^{(2)}$ & $(4.5^{+1.7}_{-1.0})\cdot 10^7\,$, $(5.5\pm 3.0)\cdot 10^7$ $^{(3)}$  \\ 
%	& & \\
	 & & \\
	Gravitational radius $R_G$            &  $4.72 \cdot 10^{14}$cm$=1.53\cdot10^{-4}$pc$^{(*)}$,     &  $7.375 \cdot 10^{12}$cm = $2.390\cdot10^{-6}\,$pc \\
	  & $9.44 \cdot 10^{14}$cm$=3.06\cdot10^{-4}$pc$^{(**)}$   & \\
 	 & & \\
	distance from Earth $d$  &   $(16.7\pm 0.2)\,$Mpc$^{(4)}$   &  $(3.8\pm0.1)\,$Mpc$^{(5)}$       \\
     & & \\	
	inclination $i$   & $\ang{(15-25)}^{(6)}$, $\ang{(30-35)}^{(7)}$,   & $\ang{(15-80)}^{(9)}$,\\
	 & $\ang{(30-45)}^{(8)}$  & $\ang{(50-80)}^{(10)}$ (pc-scale)  \\
	 & & \\
    nuclear X-ray luminosity $L_X$  &  $7.0\cdot 10^{40}$ erg$\cdot$s$^{-1}$ (($0.5-7$)keV)$^{(11)}$   & $\sim 5 \cdot 10^{41}$ erg$\cdot$s$^{-1}$ (($2-10$)keV)$^{(12)}$  \\
	 & & \\
	nuclear bolometric & $ \sim 10^{42}$erg s${^{-1}}^{(13)}$ & $\sim 10^{43}$erg s${^{-1}}^{(14)}$\\
	luminosity $L_{bol}$   & & \\
	 & & \\
 	  \hline
    \end{tabular}
   \\ $^{(*)}$ for $m=3.2\cdot 10^9$,  $^{(**)}$ for $m=6.4\cdot 10^9$,  (1) \cite{Macchetto97}, (2) \cite{GebhardtThomas09}, (3) \cite{Neumayer10}, (4) \cite{Mei07}, (5) \cite{HarrisRejkubaHarris09} , (6) \cite{Acciari09},(7) \cite{BicknellBegelman96}, (8) \cite{LyWalkerJunor07}, (9) \cite{Aharonian09} and references therein,  (10)\cite{Tingay98}, (11) \cite{DiMatteo03}, (12)\cite{Evans04}, (13) \cite{Reynolds96} , (14) \cite{Karovska02}
    
    \caption{Features of the LLAGNs M87 and Cen\,A}
    \label{tablefeatures}
\end{table}
%% end \section{Introduction}
%%%%%%%%%%%%%%%%%%%%%%%%%%%%%%%%%%%%%%%%%%%%%%%%%  
%%%%%%%%%%%%%%%%%%%%%%%%%%%%%%%%%%%%%%%%%%%%%%%%%  

\section{Accretion in LLAGNs}
Currently the reason for the difference in luminosity between luminous AGNs and LLAGNs is not clear. One possibility is that their accretion modes are different. The mode of accretion depends on the amount of available accretion material and how efficiently the gravitational energy of accreted material is converted into electromagnetic radiation.
In black hole accretion mechanics the gravitational potential energy converted to radiation energy produces an associated luminosity $L_{acc}=\eta \dot{M} c^2$. Here, $\dot{M}$ is the mass accretion rate and the factor $\eta$ measures the efficiency of gravitational-energy-to-radiation conversion. $\eta=1$ implies total conversion of the released gravitational energy into radiation, but in-fallen particles can retain some of their kinetic energy, meaning that not all of the energy is converted into radiation energy and hence $\eta$ is below unity. After crossing the event horizon of the black hole at the Schwarzschild radius, $R_S=2R_G$, where
\begin{equation}
R_G=\frac{GM_{\odot}}{c^2} \ m \approx  1.475 \cdot 10^5 \  m \ \text{cm}
\end{equation}
is the gravitational radius, the kinetic energy carried with the material into the black hole contributes to its mass. For accretion onto a black hole, $\eta \sim 0.1$ is a reasonable and well-used estimate \citep{FrankKingRaine92}.

Of great importance in accretion physics is the Eddington luminosity. It is the released electromagnetic radiation energy in spherical symmetric accretion assuming hydrostatic equilibrium,
\begin{equation}
     L_{Edd} = \frac{4 \pi c G M \mu m_p}{\sigma_T} \ = \ \frac{4 \pi c G M_{\odot} \mu m_p}{\sigma_T} \ m \ \approx 1.5 \cdot 10^{38}  \ m \ \text{erg s}^{-1}  \ \propto m ,
\end{equation}
with the mean molecular weight $\mu$, the proton rest mass $m_p$ and the Thomson cross section $\sigma_T$.
The Eddington accretion rate $\dot{M}_{Edd}$ is defined as the accretion rate required to achieve $L_{Edd}$, $L_{Edd}= \eta \dot{M}_{Edd} c^2$. For the mass accretion rate in units of the Eddington accretion rate it follows that
\begin{equation}
    \dot{m}_{Edd} := \frac{\dot{M}}{\dot{M}_{Edd}} = \frac{\eta \dot{M} c^2}{\eta \dot{M}_{Edd} c^2} = \frac{L_{acc}}{L_{Edd}}   \approx \frac{L_{acc}}{1.5 \cdot 10^{38} \text{erg}\,\text{s}^{-1} \ m}    \propto \frac{L_{acc}}{m} ,
\end{equation}
and with this the ratio of the accretion rate to the black hole mass
\begin{equation}
\frac{\dot{m}_{Edd}}{m} \propto \frac{L_{acc}}{m^2}.
\end{equation}
Several astrophysical sources are observed to emit at the order of $L_{Edd}$. \\

The accretion mode in luminous AGNs must be radiatively efficient, making the Standard Thin Accretion Disk (SAD) or Shakura-Sunyaev Disk \citep[SSD,][]{ShakuraSunyaev} the favored accretion model. The SSD has a mass accretion rate at or slightly below the Eddington limit ($\dot{m}_{Edd} \lesssim 1$) and provides high optical depth. 

The accretion mode of an LLAGN is still unclear. It might proceed in a radiatively inefficient way. There exist several models for such Radiatively Inefficient Accretion Flows \citep[RIAFs, ][]{Yuan03} which offer differing reasons for radiation inefficiency. 

One way to describe AGN accretion is that during their lifetime they switch between two phases. There is a short active phase with a high mass accretion rate near the Eddington accretion rate ($\dot{m}_{Edd} \lesssim 1 $) demonstrating high radiation efficiency ($\eta \approx 0.1$), as well as a long quiescent stage, during which the mass accretion rate lies a few orders of magnitude below $\dot{M}_{Edd}$ ($\dot{m}_{Edd} \ll 1 $) and the radiation efficiency is relatively low ($\eta \ll 1$). Therefore, luminous and low luminosity AGNs seem to represent two different stages in the lives of supermassive black holes in galactic centers. While AGN black holes seem to spend most of their lifetime in a quiescent stage, there is strong evidence that their most substantial growth primarily occurred during their brief active period \citep[][and references therein]{Maoz07}.

Even though observations show evidence for transitions from such low- to high-accretion rate phases, the existence of a RIAF has not yet been proven. Furthermore, \cite{Maoz07} has analyzed the SEDs of 13 nearby galaxies with low-ionization nuclear emission-line region (LINER) nuclei, the most common form of LLAGNs. As a type 2-LINER, M87 was among the studied objects. The result of the study was that, contrary to common opinion, the objects exhibit significant nonstellar UV flux. Indeed, the UV/X-ray luminosity ratios are similar to those of Seyfert 1 nuclei, which have $\sim 10^4$ times higher luminosities. Nevertheless, the radio emission of these objects is remarkable. The luminosity ratio between radio and other wavelengths is heavily dependent on the luminosity of the object and increases with decreasing luminosity. Hence it is possible that LLAGNs (at least partially) accrete in form of a thin SSD. Therefore, as a first test, we assume an SSD as the accretion flow model in the LLAGNs M87 and Cen\,A to calculate the absorption of TeV photons produced near the black hole by SSD photons.
\subsection{Shakura-Sunyaev Disk (SSD)}  
In order to fall into the supermassive black hole material inside the accretion disk must lose some of its angular momentum. Because the total angular momentum of the disk is conserved, the angular momentum of the infalling material must be transferred to the remaining material inside the disk. In the SSD the Magnetorotational Instability \citep[MRI,][]{BalbusHawley91} is believed to be the mechanism responsible for this process. In consequence of the gain of angular momentum the remaining material moves in the opposite direction (away from the black hole). This implies that not the entirety of accretion material can actually be accreted onto the black hole. Material with non-zero angular momentum can only fall into the black whole if there is other material that can move away from the innermost regions.

The SSD is the best model to describe accretion in luminous AGNs because it is the accretion mode with the highest efficiency of gravitational energy dissipation, leading to very luminous disks. The gravitational energy of the accreted material is at first converted into kinetic energy while the material moves towards the black hole and then by viscous stresses into thermal energy, which can then be radiated away. Because the radiative cooling is very efficient, the material becomes relatively cool, and the accretion flow assumes a geometrically thin, disk-like shape. The luminosity of the whole disk lies in the range of (if not a little below) the Eddington luminosity.

Two further important properties of the SSD are the inner and outer radii, $r_{min}=R_{min}/R_G$ and $r_{max}=R_{max}/R_G$ respectively. The inner radius is well defined and lies at $r_{min}=R_{ISCO}/R_G=6$, which is the innermost stable circular orbit around a Schwarzschild black hole. At smaller distances the material falls directly into the black hole. The outer radius is not well defined. It can, however, be estimated that for a Schwarzschild black hole $50\%$ of the energy is emitted between $r=6$ and $r=30$ \citep{Netzer06}. This implies that the highest amount of radiation is emitted in the inner part of the SSD, and therefore $r_{max} \approx 2000$ is a good approximation for the outer radius of an SSD.\\

\begin{figure}
\centering{
\includegraphics[width=16.cm]{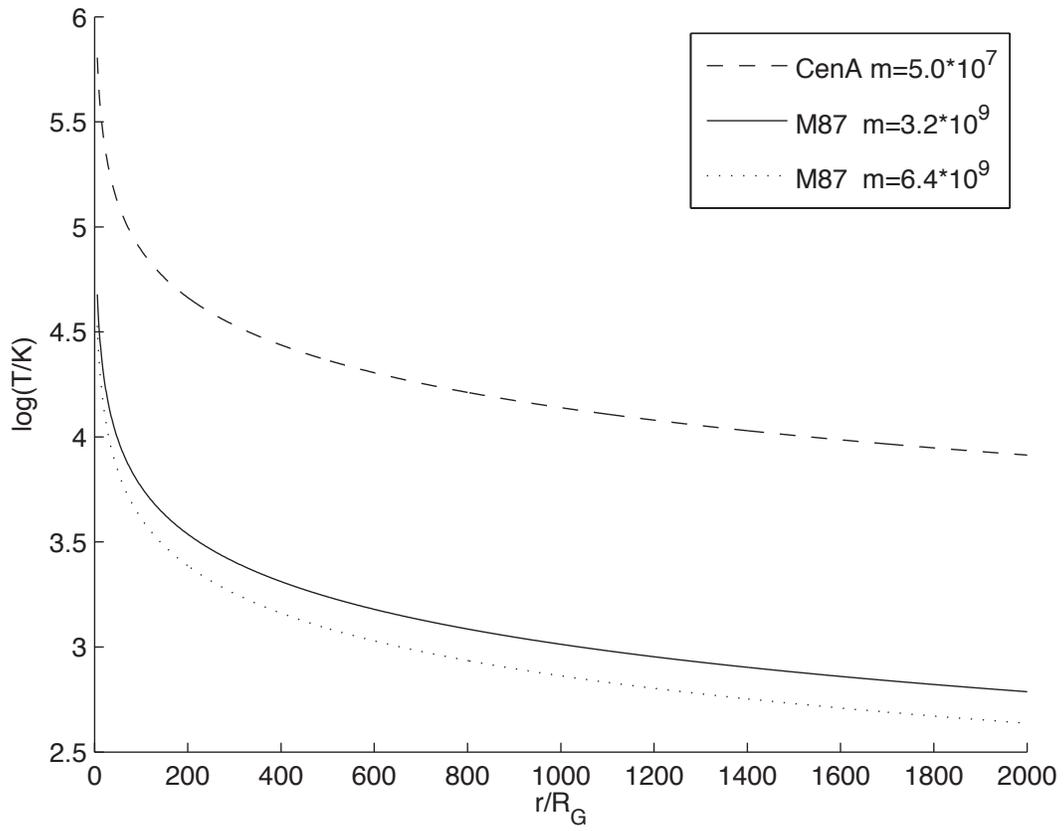}
\caption{\label{plot_temperature} Temperature profile for M87 and Cen\,A.}}
\end{figure}

The material inside the SSD is assumed to be in local thermal equilibrium and to radiate like a perfect blackbody. The temperature at a radius $r = R/R_G$ is then
\begin{align}
\Theta(r) \ & = \ \frac{k_B \, T(r)}{m_e c^2}\   = \ 1.853 \cdot 10^{-2} \  \biggl ( \frac{\dot{m}_{Edd}}{m \, r^3} \biggr)^{\frac{1}{4}} \ \biggl( 1 - \sqrt{\frac{6}{r}} \biggl )^{\frac{1}{4}} \ \\ & = \ \frac{\dot{m}_{Edd}}{m} \ Y(r) \ \ = \ \frac{L_{acc}}{m^2} \ \frac{Y(r)}{1.5 \cdot 10^{38} \text{erg}\,\text{s}^{-1} }  = \ \frac{L_{acc}}{m^2} \ \tilde{Y}(r) \propto \frac{L_{acc}}{m^2} ,
\end{align}  
where $k_B$ is the Boltzmann constant, and the function
\begin{equation}
\tilde{Y}(r) :=  \frac{Y(r)}{1.5 \cdot 10^{38} \text{erg}\,\text{s}^{-1} }  \text{\hspace*{.8cm} with \hspace*{.8cm} } Y(r) \ := \ 1.853 \cdot 10^{-2} \ r^{-3/4} \  \ \biggl( 1 - \sqrt{\frac{6}{r}} \biggl )^{\frac{1}{4}}
\end{equation}
contains the radial dependence. The highest temperature is achieved near the inner edge of the SSD. Figure \ref{plot_temperature} shows the profile of the temperatures of the SSDs in M87 and Cen\,A. The absolute temperature only depends on the ratio of the disk luminosity, $L_{acc}$, and the square of the black hole mass, $m$. The spectrum of the whole SSD is the superposition of many blackbody spectra from SSD regions with different temperatures.

\subsection{Low-luminosity SSD}
Since the luminosity of LLAGNs lies orders of magnitude below their Eddington luminosity, we set the observed bolometric luminosity $L_{bol}$ to be the luminosity of the SSD released due to accretion, $L_{bol}=L_{acc}$. The bolometric luminosity of the nucleus of M87 lies in the range of $L^{M87}_{bol} \sim 10^{42} \,$erg s$^{-1}$ \citep{Reynolds96}. This implies a mass accretion rate of $ \dot{m}^{M87}_{Edd} \sim 10^{-6}$. Although the nucleus of Cen\,A was observed at an only slightly higher luminosity of $L^{Cen\,A}_{bol} \sim 10^{43} \,$erg s$^{-1}$ \citep{Karovska02}, due to the black holes mass being two orders of magnitude reduced from the mass of the black hole in M87, the mass accretion rate in Cen\,A lies three orders of magnitude above the one in M87 ($ \dot{m}^{Cen\,A}_{Edd} \sim 10^{-3}$). Furthermore, it follows for the ratio of disk luminosity and the square of the black hole mass, that
\begin{align}
\frac{L_{bol}}{m^2} \approx
\begin{cases}
    \  10^{23}\,\text{erg\,s$^{-1}$} & \text{for M87 and $m=3.2 \cdot 10^9$}\\
   \   2.5 \cdot 10^{22}\,\text{erg\,s$^{-1}$} & \text{for M87 and $m=6.4 \cdot 10^9$} \\
    \   4 \cdot 10^{27}\,\text{erg\,s$^{-1}$} & \text{for Cen\,A and $m=5.0 \cdot 10^7$} 
  \end{cases}
  \label{Lbolm2}
\end{align}
In the following chapter we use these values to calculate the optical depth for TeV photons produced near the black hole.

\section{Attenuation of TeV photons by the SSD radiation field}
In astrophysics electron-positron pair production in photon-photon collisions is of utmost importance for high-energetic photons crossing low-energy photon fields and can lead to a significant absorption of the former. Aforementioned, we consider this process as the main source of the attenuation of TeV photons in M87 and Cen\,A.

\subsection{Electron-positron pair production in photon-photon collisions}
If the energy of the two photons (with energies $\epsilon_1=E_1/m_ec^2$ and $\epsilon_2=E_2/m_ec^2$) in their center-of-mass (COM) system is at least as large as the rest energy of two electrons then the photons can interact and create an electron-positron pair. The total cross section for this process is \citep{GouldSchreder67}:
\begin{equation}
	\sigma_{\gamma \gamma}(\beta_{cm}) \   = \  \frac{3}{16} \sigma_T (1-\beta_{cm}^2) \Biggl[ (3 - \beta_{cm}^4) \cdot  \ln{\biggl( 		\frac{1 + \beta_{cm}}{1 - \beta_{cm}} \biggr) } - \ 2 \beta_{cm} (2 - \beta_{cm}^2) \Biggr] \ := \ \frac{3}{16} \, \sigma_T \ \tilde{\sigma}_{\gamma \gamma}(\beta_{cm}),
\end{equation}
where  $\sigma_T$ is the Thomson cross-section for an electron, $ \beta_{cm} \ = \ \sqrt{1 - \gamma^{-2}_{cm}}  \ = \ \sqrt{1 - 2/(\epsilon_{\gamma}\epsilon_d(1-\mu))}$ is the electron/positron velocity in the COM system in units of the speed of light and $\gamma_{cm}$ is the electron/positron COM frame Lorentz factor, respectively. The strength of this collision is characterized by the invariant energy $\epsilon_{tot} = E_{tot}/m_e c^2$:
\begin{equation}
\epsilon^2_{tot} \, = \, 2 \epsilon_1 \epsilon_2 (1-\mu),
\label{Etot}
\end{equation}
where  $ \mu = \cos{\theta} $ is the Cosine of the angle $\theta$ between the interacting photons in the laboratory system. To create an electron/positron pair (each of them with energy $E_{e^{\pm}} = \gamma_{cm} m_e c^2 $), the total energy has to be
\begin{align}
E_{tot} = 2 \gamma_{cm} m_e c^2 \ \ \text{\hspace*{.5cm}or \hspace*{.5cm} } \ \ &  \epsilon_{tot} = 2 \gamma_{cm}.
\end{align}
From this and eq.(\ref{Etot}) it follows that
\begin{align}
\epsilon^2_{tot} = \, 2 \epsilon_1 \epsilon_2 (1-\mu) = 4 \gamma^2_{cm} \ \ \text{\hspace*{.5cm} $\Rightarrow$ \hspace*{.5cm} } \ \ & \epsilon_1 \epsilon_2 (1-\mu) = 2 \gamma^2_{cm}.
\end{align}
The highest probability for this interaction lies near the threshold energy, where $\gamma_{cm}=1$,
\begin{equation}
\epsilon_1 \epsilon_2 (1-\mu) = 2.
\label{threshold}
\end{equation}
Therefore, the highest probability of interaction of TeV photons with disk photons occurs at energies around
\begin{equation}
E_{disk} = \epsilon_{disk} m_e c^2 = \frac{2 (m_ec^2)^2}{E_{\gamma} \, (1-\mu)} \approx \frac{0.5 \text{eV}}{ \frac{E_{\gamma}}{\text{TeV}}  (1-\mu)} \sim \text{eV}.
\label{IR}
\end{equation}
%xxxxxxxxxxxxxxxxxxxxxxxxxxxxxxxxxxxxxxxxxxxxxxxxxxxxxxxxxxxxxxxxxxxxxxxxxxxxxxxxxxxxxxxxxxxxxxxxxxxxxxxxxxxxxxxxxxxxx
\subsection{Optical Depth}
Determining the absorption probability for the TeV photon (with energy $\epsilon_{\gamma}$) traversing the disk photon field (photon energy $\epsilon_d$, photon density $n_d(\epsilon_d)$) is achieved by calculating the optical depth for this process \citep{JauchRohrlich}:
\begin{equation}
	\tau_{\gamma\gamma}(\epsilon_{\gamma}, l) \  = \ \int\limits_0^{l} dl \ \oint d\Omega \, (1-\mu) \ \int\limits_{2\,/\,(\epsilon_{\gamma}\, (1-\mu))}^{\infty} d\epsilon_d \ \sigma_{\gamma \gamma}(\epsilon_{\gamma},\,\epsilon_d ,\, \mu) \cdot n_d(\epsilon_d).
\end{equation}
TeV radiation with the initial intensity $I_0$ is attenuated to the value
\begin{equation}
I(l,\epsilon_{\gamma})\ = \ I_0(\epsilon_{\gamma}) \, e^{-\tau(\epsilon_{\gamma}, l)}
\end{equation}
after the travel distance $l$ through the absorbing medium. For $\tau \ll 1$ there is no strong absorption, and $I(l) \approx I_0$, while in the case of $\tau \geq 1$ the TeV radiation is attenuated.
%xxxxxxxxxxxxxxxxxxxxxxxxxxxxxxxxxxxxxxxxxxxxxxxxxxxxxxxxxxxxxxxxxxxxxxxxxxxxxxxxxxxxxxxxxxxxxxxxxxxxxxxxxxxxxxxxxxxxx
\subsection{Internal $\gamma\gamma$-opacity in LLAGNs}

\begin{figure}
\centering{
\includegraphics[width=14.cm]{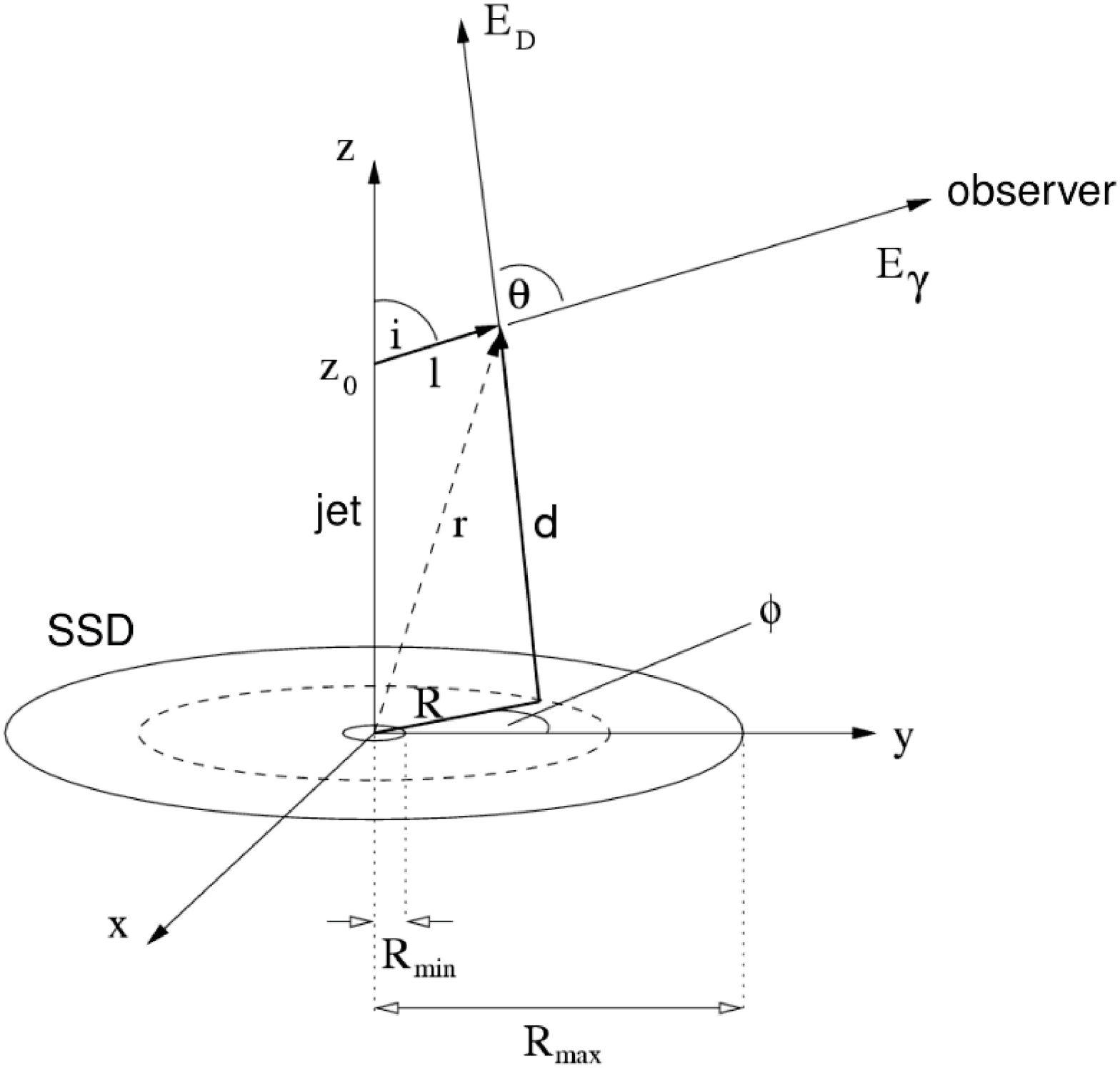}
\caption{\label{Modellskizze} Schematic representation of the inner part of an LLAGN.}}
\end{figure}

Figure \ref{Modellskizze} shows the inner regions of an LLAGN with the supermassive black hole located in the point of origin. The SSD lies in the x-y-plane and perpendicular to the disk is the propogating jet along the z-axis. At a height $z_0$ above the black hole the TeV photon is produced and is able to interact with one of the SSD photons. We assume the SSD extends from the radius $r_{min} = 6$ to the radius $r_{max} = 2000$, as noted previously.

%____________________________________________________________________________________________________________
Every SSD area $dA\, = \, dr\ r \ d\phi$ produces disk photons at the interaction point with spectral photon density
\begin{align}
\begin{split}
n(\epsilon_d) \ & = \ \frac{15}{4 \pi^5 m_e c^3} \ \frac{dL_{acc}}{dA} \  \frac{\epsilon_d^2}{\Theta^4 (e^{\epsilon_d/ \Theta} - 1)}  \ \frac{\delta(z)}{d^2} \\
& = \ \frac{15}{4 \pi^5 m_e c^3} \ \frac{L_{acc}}{\pi R^4_G (r_{max}^2 - r_{min}^2)} \ \frac{\epsilon_d^2}{\Theta^4 (e^{\epsilon_d/ \Theta} - 1)}  \ \frac{\delta(z)}{d^2}.
\end{split}
\end{align} 
With the assumption that the energy output is the same for every part of the SSD,
\begin{equation}
\frac{dL_{acc}}{dA} \ = \ \frac{L_{acc}}{\pi \, R_G^2\, (r_{max}^2 - r_{min}^2)}.
\end{equation}
Here, $d$ is the distance between the emission site of the SSD photon and the $\gamma$ interaction point \citep{BoettcherDermer05},
\begin{equation}
d^2 = r^2 R_G^2 + l^2 + z_0^2 + 2\,l\,(z_0 \cdot \cos{i} - r \, R_G \cdot \sin{i} \cdot \cos{\phi}) \ ,
\end{equation}
and 
\begin{equation}
\mu = \frac{l + z_{0} \cdot \cos{i} - r \, R_G \cdot \sin{i} \cdot \cos{\phi}}{d} \ .
\end{equation}
Therefore, the optical depth is 
\begin{align}
\begin{split}
\tau_{\gamma\gamma}(\epsilon_{\gamma}) \ & =        \frac{45}{16\pi^5 (r^2_{max}-r^2_{min})}               \frac{m_p}{m_e}    \ \      \dot{m}_{Edd}          \ \int\limits_0^{\infty} dl \             \int\limits_{r_{min}}^{r_{max}} dr \frac{r}{\Theta^4(r)}                          \int\limits_{0}^{2\pi} d\phi                   \, \frac{1-\mu}{d^2} \     \int\limits_{\frac{2}{\epsilon_{\gamma}\,	(1-\mu)}}^{\infty} d\epsilon_d \ \frac{ \epsilon^2_d \ \tilde{\sigma}_{\gamma \gamma}(\epsilon_{\gamma}, \,\epsilon_d , \,\mu) }{\exp{ \bigl ( \epsilon_d / \Theta(r) \bigr )  }-1} \\
& =  \frac{45}{16\pi^5 (r^2_{max}-r^2_{min})}               \frac{m_p}{m_e}   \ \  m       \ \int\limits_0^{\infty} dl \             \int\limits_{r_{min}}^{r_{max}} dr \frac{r}{Y^4(r)}              \int\limits_{0}^{2\pi} d\phi                   \, \frac{1-\mu}{d^2} \                 \int\limits_{\frac{2}{\epsilon_{\gamma}\,	(1-\mu)}}^{\infty} d\epsilon_d \ \frac{ \epsilon^2_d \ \tilde{\sigma}_{\gamma \gamma}(\epsilon_{\gamma}, \,\epsilon_d ,\,\mu)}{\exp{ \Bigl ( \frac{m^2}{L_{bol} }   \frac{\epsilon_d}{\tilde{Y}(r)} \Bigr )  }  -1} \ \ .
\label{opticaldepth}
\end{split}
\end{align}
As can be seen from this expression, the optical depth for the TeV photons depends on the black hole mass, the disk luminosity, and the inclination of the AGN. In the following we discuss the expected influence of these quantities on the optical depth.

\subsubsection{Inclination $i$}
The inclination of the AGN determines (1) the distance between the TeV photon and the SSD and therefore the density of disk photons in the trajectory of the TeV photon and (2) the angle between the scattering photons. An increase in the inclination implies a higher SSD photon density and hence a higher optical depth, however the influence of the latter effect on optical depth is not as obvious.
Since the probability of interaction is highest near the threshold of electron-positron pair production (cf. eq.(\ref{threshold})), the highest probability of interaction between any given TeV photon and a disk photon depends highly on the interaction angle. 
For low inclinations ($i < \ang{45}$) interaction angles of $\theta>\pi/2$ are rather unlikely, whereas for larger inclinations ($\ang{45} < i < \ang{90}$) interaction angles $\theta<\pi/2$ are more probable. Therefore, TeV photons emitted at low inclinations interact with higher energy disk photons than TeV photons emitted under a larger angle to the jet. Therefore, whether an increased $i$ increases or decreases the optical depth depends on the SSD spectrum.

Furthermore, for very small inclinations $i \lesssim \ang{10}$ relativistic effects must be considered. In such cases the particles inside the jet move with highly relativistic velocities along our line of sight, therefore the emitted radiation is observed at higher energy and intensity than it had inside the source. Due to this relativistic beaming, blazars can be observed on Earth up to TeV energies even though the emitting particles did not produce such high energetic radiation in their COM system.

\subsubsection{Black hole mass $m$}
The mass of the black hole greatly affects the physics of the accretion disk so intuitively one can predict it will vary the optical depth calculation. The temperature of the flow (and therefore the emitted blackbody spectrum), the scaling of the AGN by the gravitational radius, as well as the mass accretion rate strongly depend on the black hole mass. On the whole, the term for the optical depth is dominated by the exponential function in its denominator (cf. eq.(\ref{opticaldepth})), so a higher black hole mass implies a lower optical depth.

\subsubsection{Bolometric luminosity $L_{bol}$ of the accretion disk}
\label{Lbol}
A decrease/increase of $L_{bol}$ and an increase/decrease of the square of the black hole mass have a similar effect on the optical depth. Therefore, a higher bolometric luminosity leads to an increase of the optical depth and vice versa.

%____________________________________________________________________________________________________________
\subsection{Comparison with other models}
The optical depth for GeV photons traveling through a disk photon field has been calculated by \cite{BeckerKafatos95} for blazars in general and for 3C 279 in particular. \cite{BeckerKafatos95} assumed a two-temperature disk with a hot inner (two-temperature) region extending from $r_{min} = 6$ to $r_{max}$, where $30 \lesssim r_{max} \lesssim 100$, and a cool (single-temperature) region for $r > r_{max}$. They assumed the X-ray disk emission relevant for TeV $\gamma$-ray absorption to be produced in the hot inner region, where electrons with a nearly constant temperature $T_e \sim (10^9\,-\,10^{10})\,$K upscatter UV radiation from the cool outer region, modelled the emission by a power-law, and scaled it to the observed X-ray luminosity. A similar calculation for M87 has been made by \cite{Cheung07} and \cite{NeronovAharonian07}, who assumed an IR target photon source with linear size of $r=1$ and $r=25$, respectively. While \cite{Cheung07} could exclude the central region as a TeV emission site, \cite{NeronovAharonian07} concluded that the core region is transparent for TeV photons. 

In contrast to these authors, in our model the energy range of the target photons depends on the properties of the special object. The SSD in M87, e.g., produces radiation in the IR and optical, while in Cen$\,$A the SSD photons have higher energies from the IR up to the Middle Ultraviolet (MUV) range of the electromagnetic spectrum. Ie. the minimum and maximum energies of the disk photons in the model used in this paper are not constant, but depend on the black hole mass and the disk luminosity. Therefore, we cannot assume a spectrum with fixed beginning and end, but the spectrum arises from the properties of the particular LLAGN. In addition, our aim is to explain the physics inside an LLAGN and to understand the reasons for the difference between luminous and low luminosity AGNs.

%xxxxxxxxxxxxxxxxxxxxxxxxxxxxxxxxxxxxxxxxxxxxxxxxxxxxxxxxxxxxxxxxxxxxxxxxxxxxxxxxxxxxxxxxxxxxxxxxxxxxxxxxxxxxxxxxxxxxxxxxxxxxx
%xxxxxxxxxxxxxxxxxxxxxxxxxxxxxxxxxxxxxxxxxxxxxxxxxxxxxxxxxxxxxxxxxxxxxxxxxxxxxxxxxxxxxxxxxxxxxxxxxxxxxxxxxxxxxxxxxxxxxxxxxxxxx
%\newpage
\section{Calculation of the optical depth for $\gamma \gamma$ interactions in M87 and Cen$\,$A}
%xxxxxxxxxxxxxxxxxxxxxxxxxxxxxxxxxxxxxxxxxxxxxxxxxxxxxxxxxxxxxxxxxxxxxxxxxxxxxxxxxxxxxxxxxxxxxxxxxxxxxxxxxxxxxxxxxxxxxxxxxxxxx
\subsection{M87}
\subsubsection{Features}
M87 (NGC 4486) is a giant elliptical and perhaps the most dominant galaxy in the Virgo Cluster. Although not much larger in its linear extension than the Milky Way, due to its elliptical shape M87 contains much more stars and mass than the Galaxy. The central region of this huge radio galaxy, M87, whose counterpart at radio wavelengths is the strong radio source Virgo A, harbors a supermassive black hole with a mass of $m=(3.2-6.4) \cdot 10^9$. This black hole seems to be what triggers the observed activity of the core region, such as the impressive $2\,$kpc long jet. Discovered in 1918 by Curtis, the jet has today been resolved over an array of wavelengths (radio, optical and X-ray). The polarization of the observed electromagnetic radiation from the jet is typical of synchrotron radiation and supports the conclusion that the jet consists of relativistically moving particles transported from the nucleus to outer regions of the galaxy. The angle between the jet and our line of sight is not well known. While \cite{BicknellBegelman96} quote a jet angle of $i=\ang{(30-35)}$, \cite{LyWalkerJunor07} assume a larger value of $i=\ang{(30-45)}$ and \cite{Acciari09} a smaller value of $i=\ang{(15-25)}$. In our calculations, we will test the cases $i=\ang{20}, \ang{30}$ and $\ang{40}$. The gravitational radius of the black hole in M87 is
\begin{align}
R_G \ & = \ \frac{GM_{\odot}}{c^2} \   \approx \  4.72 \cdot 10^{14} \ \biggl ( \frac{m}{ 3.2 \cdot 10^9} \biggr ) \ \text{cm}  = \
\begin{cases}
    \ 4.72 \cdot 10^{14}\text{cm} = 1.53 \cdot 10^{-4}\text{pc} & \text{for $m=3.2 \cdot 10^9$}\\
   \  9.44 \cdot 10^{14}\text{cm} = 3.06 \cdot 10^{-4}\text{pc} & \text{for $m=6.4 \cdot 10^9$} \hspace*{.4cm} .
  \end{cases}
\end{align}

Although M87 harbors an AGN, the optical emission from the nucleus is rather low. With an observed bolometric luminosity of the nucleus of $L_{bol} \sim 10^{42}$erg s${^{-1}}$ \citep{Reynolds96}, M87 lies $5$ to $6$ orders of magnitude below its Eddington luminosity and is therefore an LLAGN. As an (LL)AGN M87 is classified as a radio galaxy of the FR I type. Furthermore M87 is a type-2 LINER galaxy, where the nucleus shows narrow emission lines but no broad line-emission. M87 has been observed at TeV energies several times \citep{Aharonian06, Albert08, Acciari08, Acciari09}. The highest measured energy of the $\gamma$-rays is $ E^{M87}_{max} \approx 20\,$TeV \citep{Acciari08}. The TeV spectrum of 2005 shows variability on timescales of days, which strongly constrains the size of the TeV $\gamma$ emission region to $\approx 5 \delta R_S$ or equivalently $10 \delta R_G$ \citep{Aharonian06}. Since the jet of M87 is misaligned, the Doppler factor is $\delta \sim 1$ \citep{Giannios10}, resulting in an extension of the emission region of $\sim 10R_G$. Moreover, radio observations of M87 from 2007 suggest that the TeV radiation is produced not more than $\sim 0.015\,$pc away from the central black hole \citep{Walker08}, which corresponds to $\sim 100\,R_G$ assuming a black hole mass of $m=3.2 \cdot 10^9$ and to $\sim 50\,R_G$ assuming a black hole mass of $m=6.4 \cdot 10^9$.

\subsubsection{Results}
The optical depth has been calculated for three different inclinations ($i=\ang{20},\,\ang{30}, \, \ang{40}$) and the two possible black hole masses ($m=3.2 \cdot 10^9$ and $m=6.4 \cdot 10^9$). In all calculations for M87 the bolometric luminosity is assumed to be $L_{bol}=10^{42}\,$erg$\,$s$^{-1}$. 

\begin{figure}
\centering{
   \includegraphics[width=16.cm]{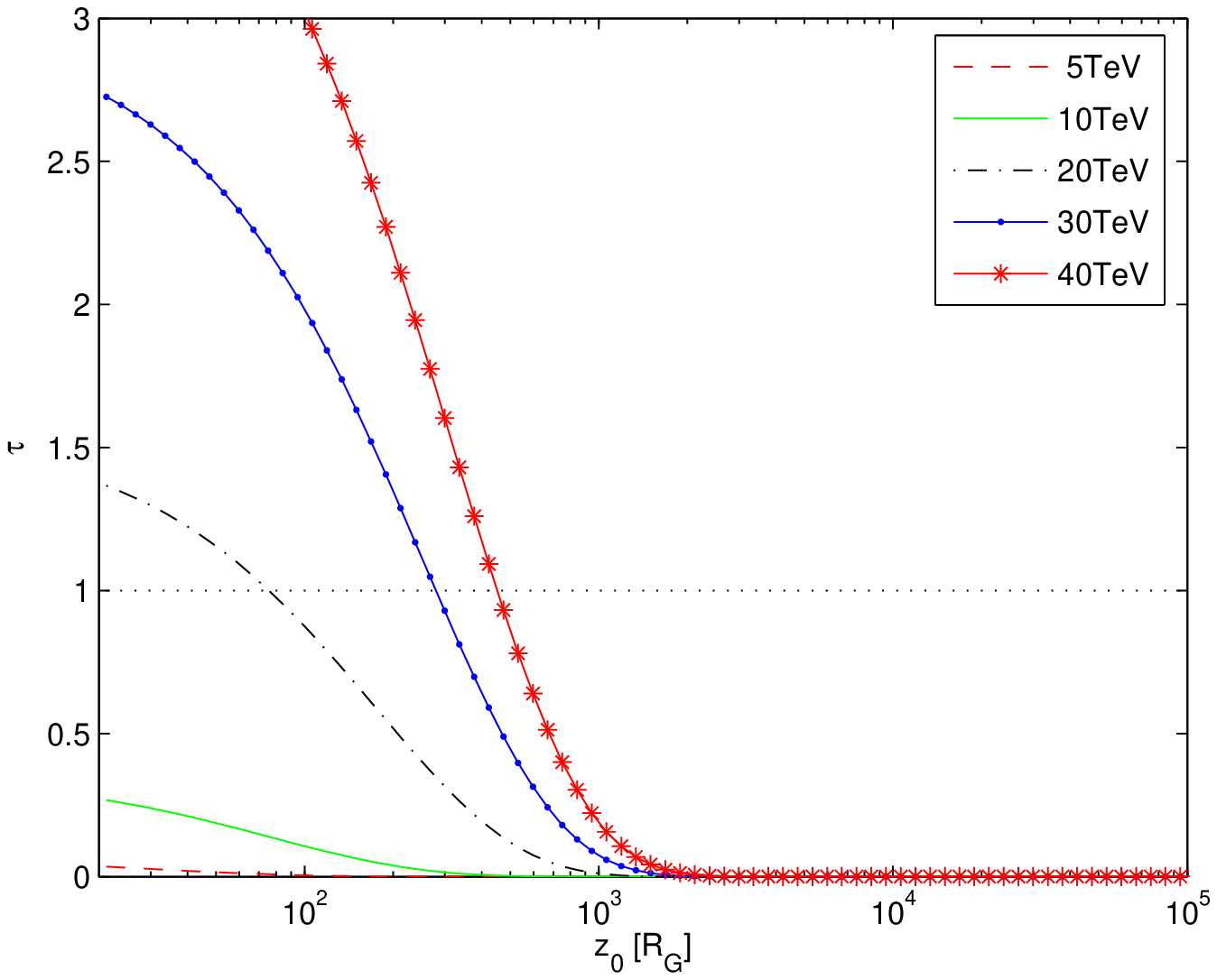} 
\caption{\label{M87_m32_i20} \small Optical depth for TeV radiation produced in M87 with $m=3.2 \cdot 10^9$ and $i=\ang{20}$.}}
\end{figure}

\begin{figure}
\centering{
  \includegraphics[width=16.cm]{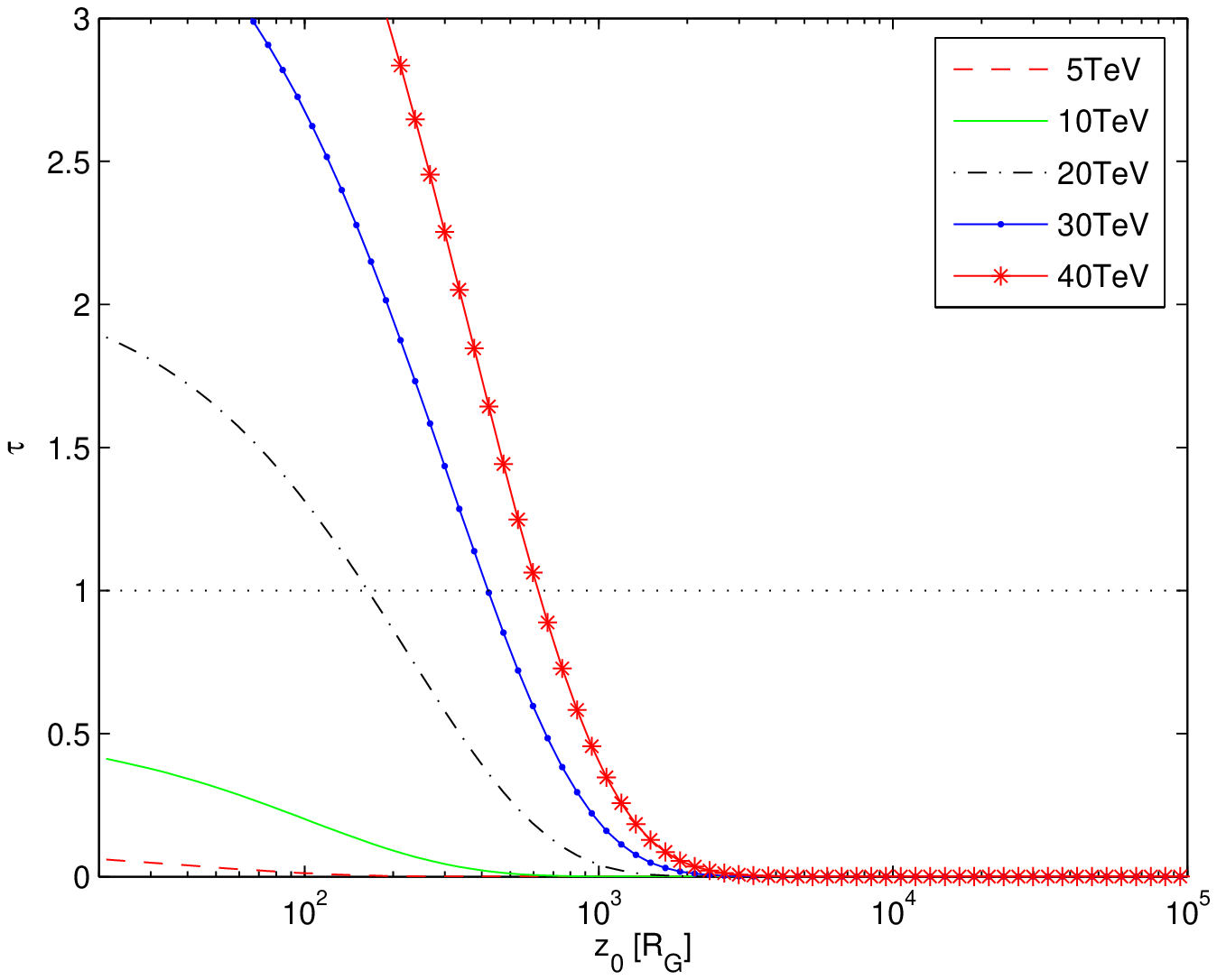} 
  \caption{\label{M87_m32_i30} \small Optical depth for TeV radiation produced in M87 with $m=3.2 \cdot 10^9$ and $i=\ang{30}$.}}
\end{figure}

\begin{figure}
\centering{
  \includegraphics[width=16.cm]{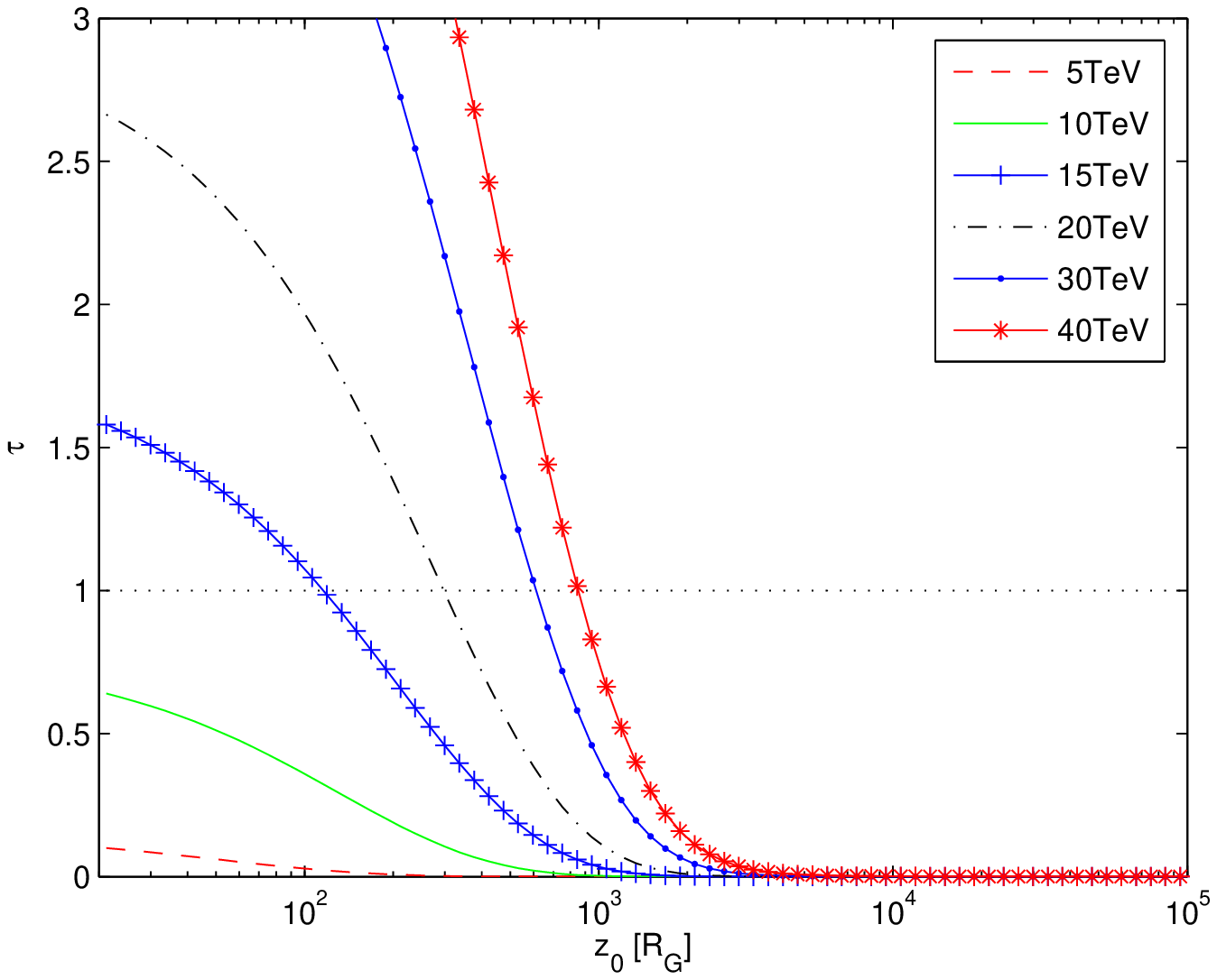} 
 \caption{\label{M87_m32_i40} \small Optical depth for TeV radiation produced in M87 with $m=3.2 \cdot 10^9$ and $i=\ang{40}$.}}
\end{figure}

In Figure \ref{M87_m32_i20} we set $m=3.2 \cdot 10^9$ and $i=\ang{20}$ and accordingly plot the optical depth $\tau_{\gamma \gamma}$ as a function of the production height $z_0$ of the TeV photon above the black hole in units of the gravitational radius. According to this value of the black hole mass, the TeV photon production should take place within $z_0\approx 100\,R_G$ above the black hole. Using this value for the production height, the optical depth exceeds unity only for photons with energy $E_{\gamma} > 20\,$TeV. For lower photon energies the optical depth is $ \ll 1$, implying no strong absorption. For $20\,$TeV photons the optical depth is slightly smaller than one, so despite attenuation there should be no total absorption, and some of these photons should be able to leave the core region. Since the highest energy of the observed TeV photons in M87 is $E_{\gamma} \approx 20\,$TeV, our results are consistent with observations. Figure \ref{M87_m32_i30} shows the optical depth for the same parameters as Figure \ref{M87_m32_i20} with the exception of a higher inclination of $i=\ang{30}$. For this inclination the optical depth is slightly higher than for $i=\ang{20}$, but the results do not change dramatically. Photons with $E_{\gamma} < 20\,$TeV should be able to escape the production region, and even $20\,$TeV photons should not be totally absorbed. %
For an even higher inclination of $i=\ang{40}$, the optical depth reaches unity for just $15\,$TeV photons and reaches the value of $\approx 2$ for $20\,$TeV photons (cf. Figure \ref{M87_m32_i40}), leading to an absorption of nearly $90\%$ of the latter. For more energetic photons the optical depth is even higher and hence the probability for escaping the SSD photon field is very small. %

\begin{figure}
\centering{
   \includegraphics[width=16.cm]{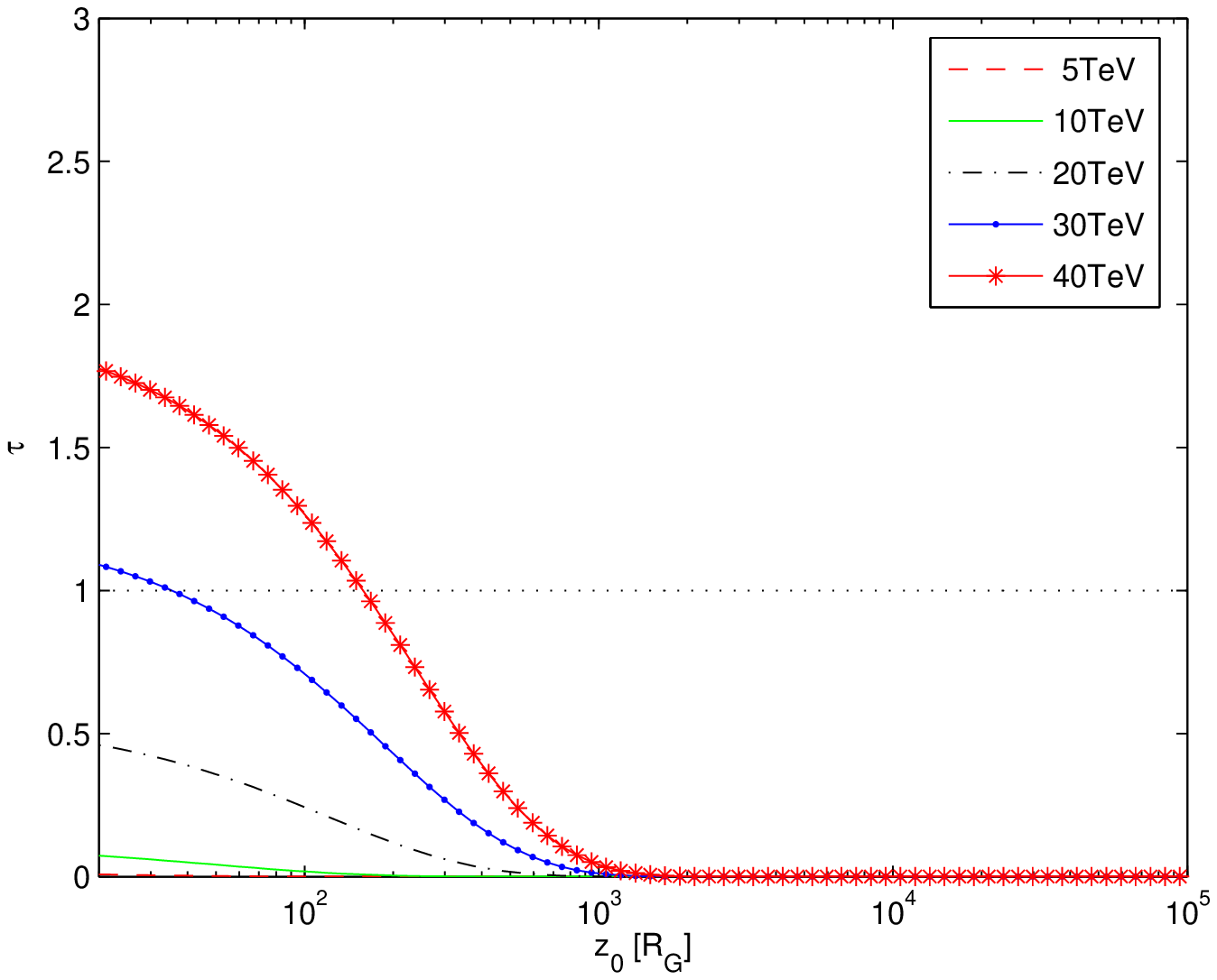} 
   \caption{\label{M87_m64_i20} \small Optical depth for TeV radiation produced in M87 with $m=6.4 \cdot 10^9$ and $i=\ang{20}$.}}
\end{figure}

\begin{figure}
\centering{
  \includegraphics[width=16.cm]{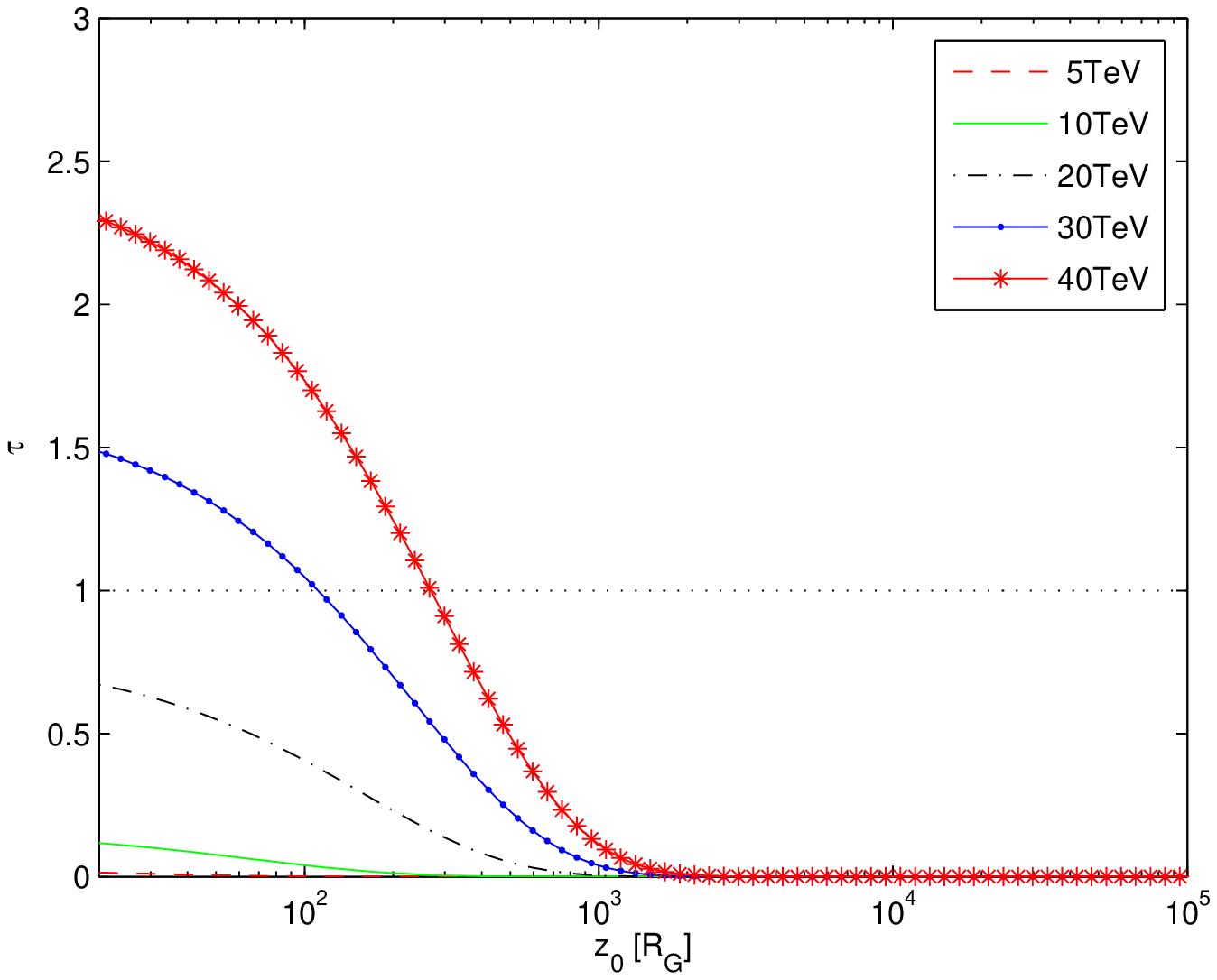} 
  \caption{\label{M87_m64_i30} \small Optical depth for TeV radiation produced in M87 with $m=6.4 \cdot 10^9$ and $i=\ang{30}$.} }
\end{figure}

\begin{figure}
\centering{
  \includegraphics[width=16.cm]{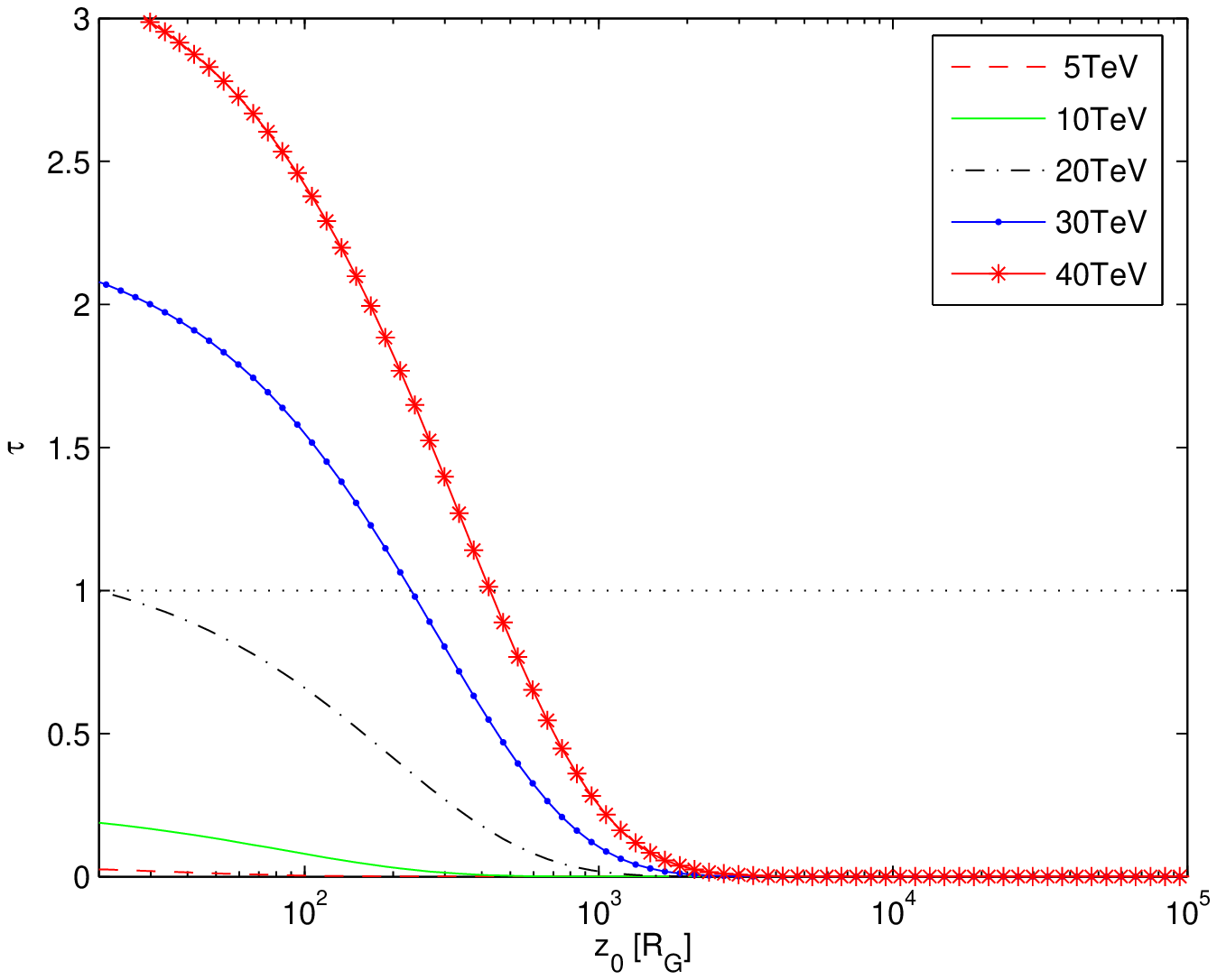} 
  \caption{\label{M87_m64_i40} \small Optical depth for TeV radiation produced in M87 with $m=6.4 \cdot 10^9$ and $i=\ang{40}$.}}
\end{figure}

If the mass of the M87 black hole has the higher value of $m=6.4 \cdot 10^9$, the central region of M87 becomes even more transparent for TeV photons, as can be seen in Figures \ref{M87_m64_i20}, \ref{M87_m64_i30} and \ref{M87_m64_i40} for an inclination of $i=\ang{20},\ \ang{30}$ and $\ang{40}$, respectively. Although in this case the emission region should lie within the innermost $\sim 50\,R_G$, the optical depth does not reach unity for the observed photon energies. For $i=\ang{20}$ photons up to $E_{\gamma} \sim 30\,$TeV should escape their production region, and despite an increase in inclination to $i=\ang{30}$ (cf. Figure \ref{M87_m64_i30}) resp. $i=\ang{40}$ (cf. Figure \ref{M87_m64_i40}) $E_{\gamma} \lesssim 20\,$TeV photons should be able to escape the core region. 

\begin{figure}
\centering{
   \includegraphics[width=16.cm]{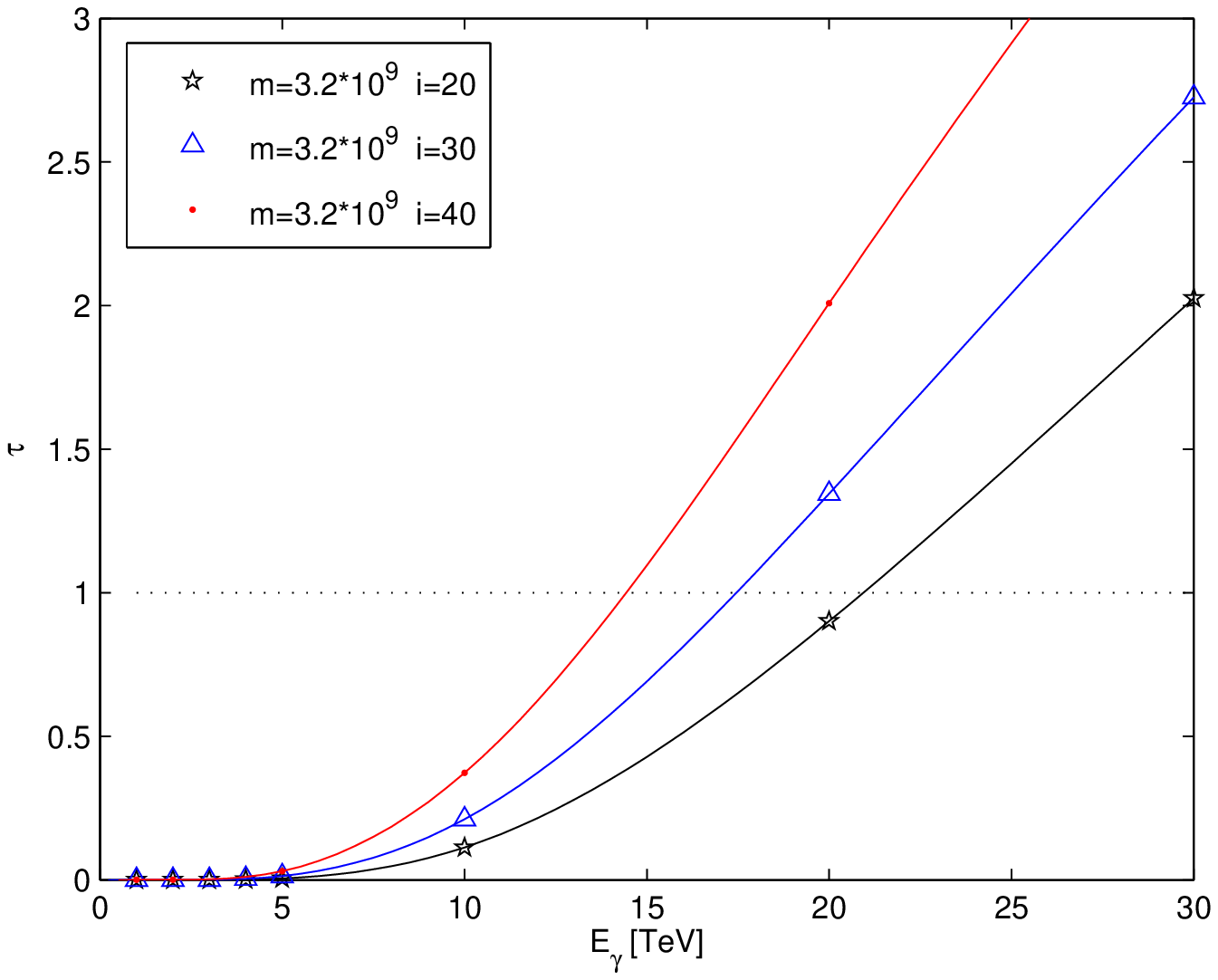} 
\caption{\label{M87_plotsenergy_100RG} \small Optical depth for TeV radiation produced in M87 for $z_0=100\, R_G$.}}
\end{figure}

\begin{figure}
\centering{
 \includegraphics[width=16.cm]{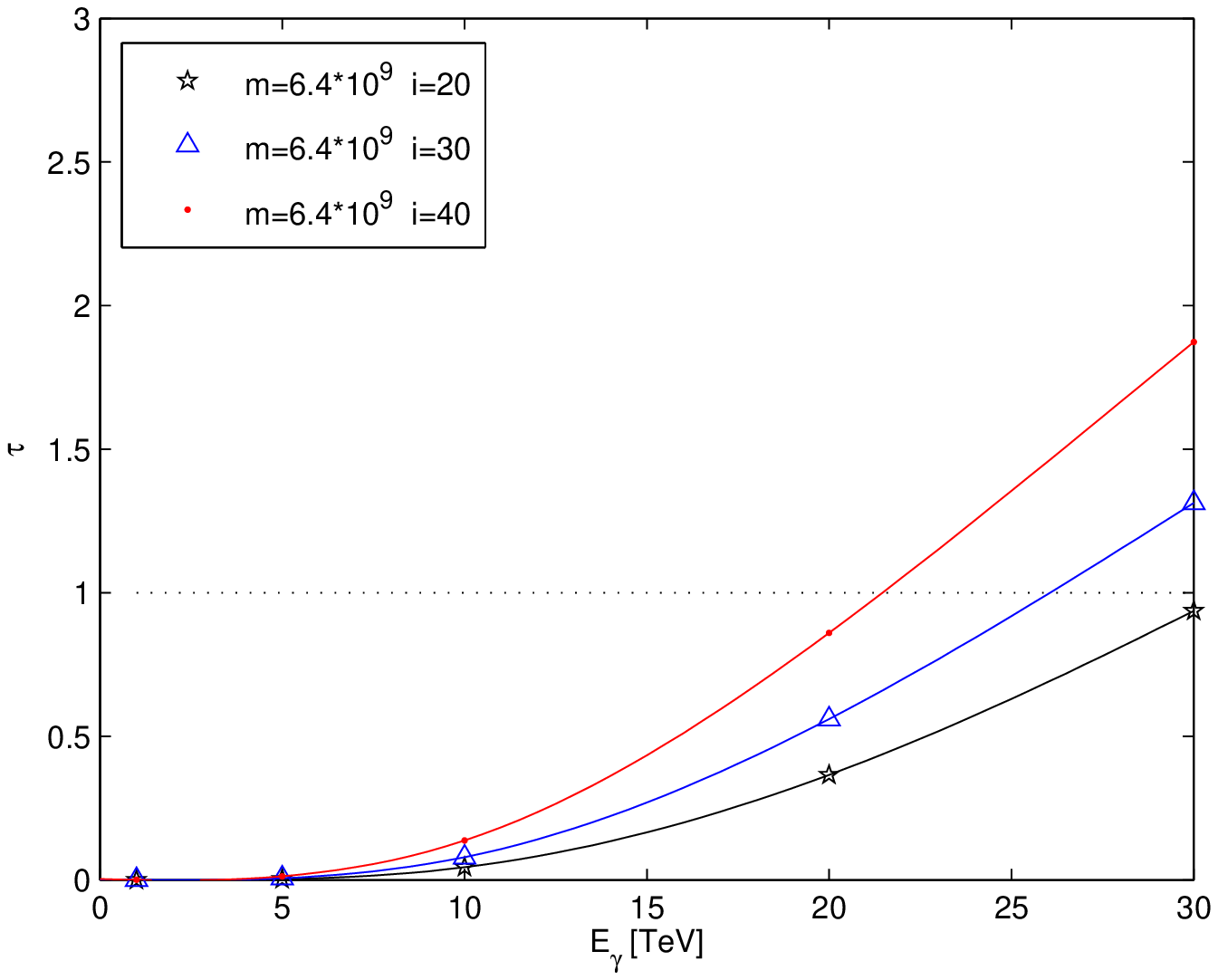} 
\caption{\label{M87_plotsenergy_50RG} \small Optical depth for TeV radiation produced in M87 for $z_0=50\, R_G$.}}
\end{figure}

To illustrate the absorption probability of the estimated production region, we plot the optical depth as a function of the TeV photon energy at the maximum possible production distance of $z_0=100\,R_G$ for $m=3.2 \cdot 10^9$ (cf.  Figure \ref{M87_plotsenergy_100RG}) and $z_0 = 50\,R_G$ for $m=6.4 \cdot 10^9$ (cf. Figure  \ref{M87_plotsenergy_50RG}). It can be seen that the optical depth for photons up to $20\,$TeV exceeds unity only for a black hole mass of $m=3.2 \cdot 10^9$ and the larger inclinations of $i=\ang{30}$ and $\ang{40}$, implying a perceptible absorption only for these parameters.

%xxxxxxxxxxxxxxxxxxxxxxxxxxxxxxxxxxxxxxxxxxxxxxxxxxxxxxxxxxxxxxxxxxxxxxxxxxxxxxxxxxxxxxxxxxxxxxxxxxxxxxxxxxxxxxxxxxxxxxxxxxxxx
%xxxxxxxxxxxxxxxxxxxxxxxxxxxxxxxxxxxxxxxxxxxxxxxxxxxxxxxxxxxxxxxxxxxxxxxxxxxxxxxxxxxxxxxxxxxxxxxxxxxxxxxxxxxxxxxxxxxxxxxxxxxxx
\subsection{Cen\,A}
\subsubsection{Features}
Centaurus A, the 5$^{\text{th}}$ brightest system visible to earthbound observers, lies in the Constellation Centaurus of the southern sky. It is a nearby, lenticular radio galaxy at a distance of $3.8\,$Mpc \citep{HarrisRejkubaHarris09}. It is the closest recent merger, a trait which could increase $\dot{M}$ and therefore, assuming the AGN is accretion powered, contribute to VHE radiation production \citep{MartiniHo04}. Cen\,A is a very interesting case study since it is one of the few VHE sources which is not a blazar \citep{Sreekumar99}, and was the first object reported by the \citet{Fermi10} to have $\gamma$-ray production clearly found coming from its massive radio lobes. Although the $\gamma$-ray flux from the lobe constitutes over half of the total flux of the galaxy, the highest energy $\gamma$-rays \citep[$ E^{Cen\,A}_{max} \approx 10\,$TeV ,][]{RaueHess10} come from the nucleus.  \\ 

Centaurus A has an X-ray flux variation timescale of a few days, and a short $\gamma$-radiation timescale. This suggests that the VHE radiation originates near the core \citep{SitarekBednarek10}. The galaxy hosts a black hole with $m = 4.5^{+1.7}_{-1.0}\cdot 10^{7}$ based on H$_2$ kinematics and $m = (5.5\pm3)\cdot 10^{7}$ based on stellar observations \citep{Neumayer10}. An approximation of $m = 5.0\cdot 10^{7}$ is used in our calculations. The inclination of the galaxy is not known with any precision, but \cite{Tingay98} quote an inclination of the inner jet of $i=\ang{(50-80)}$. Based on this estimate we calculate the optical depth for the two inclinations $i=\ang{50}$ and $i=\ang{80}$. The spectrum of Cen\,A extends from radio wavelengths up to $\gamma$-rays with $E_\gamma \approx 10\,$TeV, with which this paper is concerned. The SED of Cen\,A is double peaked, with one peak at $\sim150\,$keV due to synchrotron emission and a second peak at $\sim1$MeV \citep{Steinle10}. The gravitational radius of the Cen\,A black hole for the mass approximation used in this paper is
\begin{equation}
R_G \approx  7.375 \cdot 10^{12}\text{cm} = 2.390 \cdot 10^{-6}\text{pc} .
\end{equation}

\subsubsection{Results}
\begin{figure}
\centering{
   \includegraphics[width=16.cm]{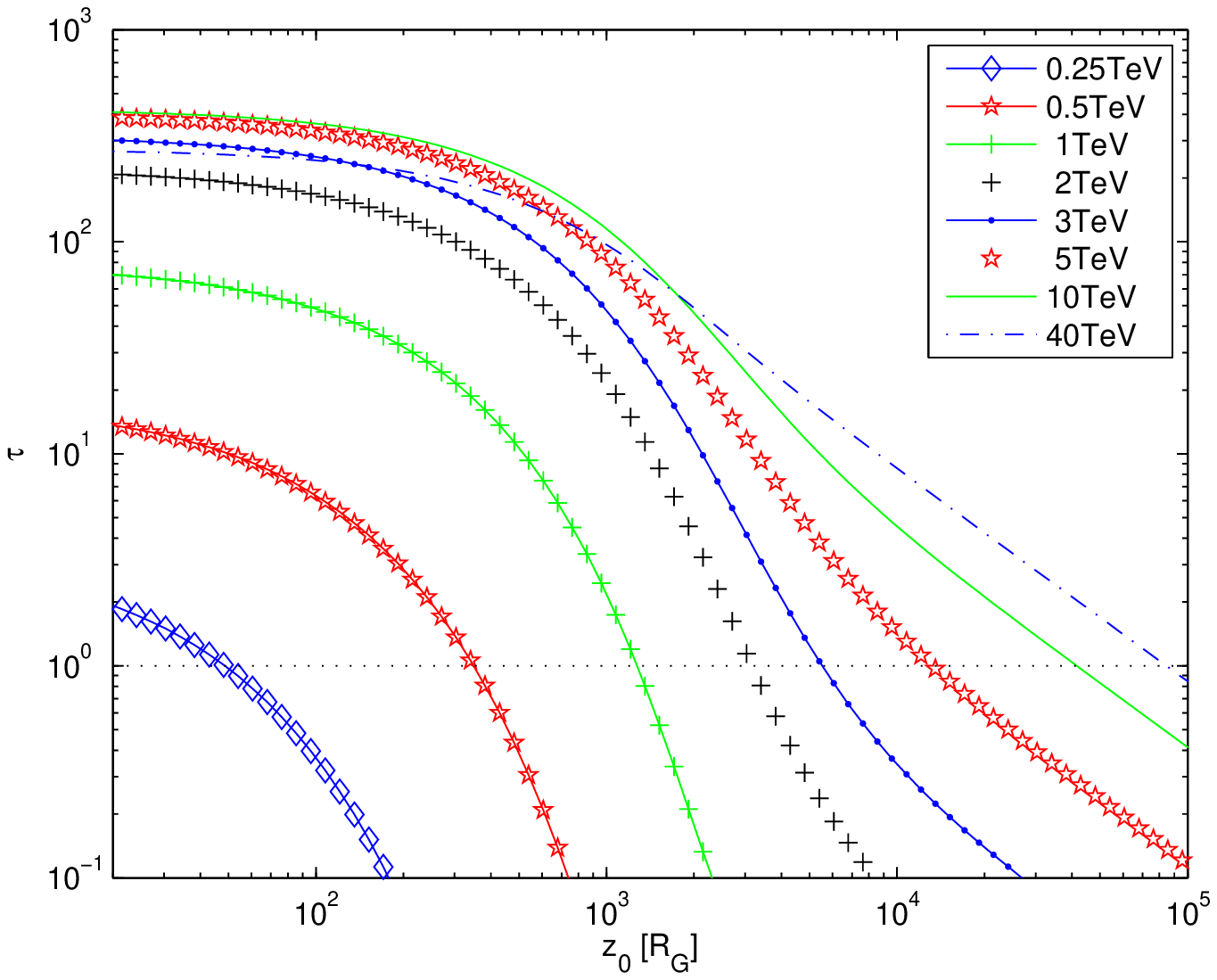}
   \caption{\label{CenA_i50} \small Optical depth for TeV radiation produced in Cen\,A with inclination $i=\ang{50}$.}}
\end{figure}

\begin{figure}
\centering{
   \includegraphics[width=16.cm]{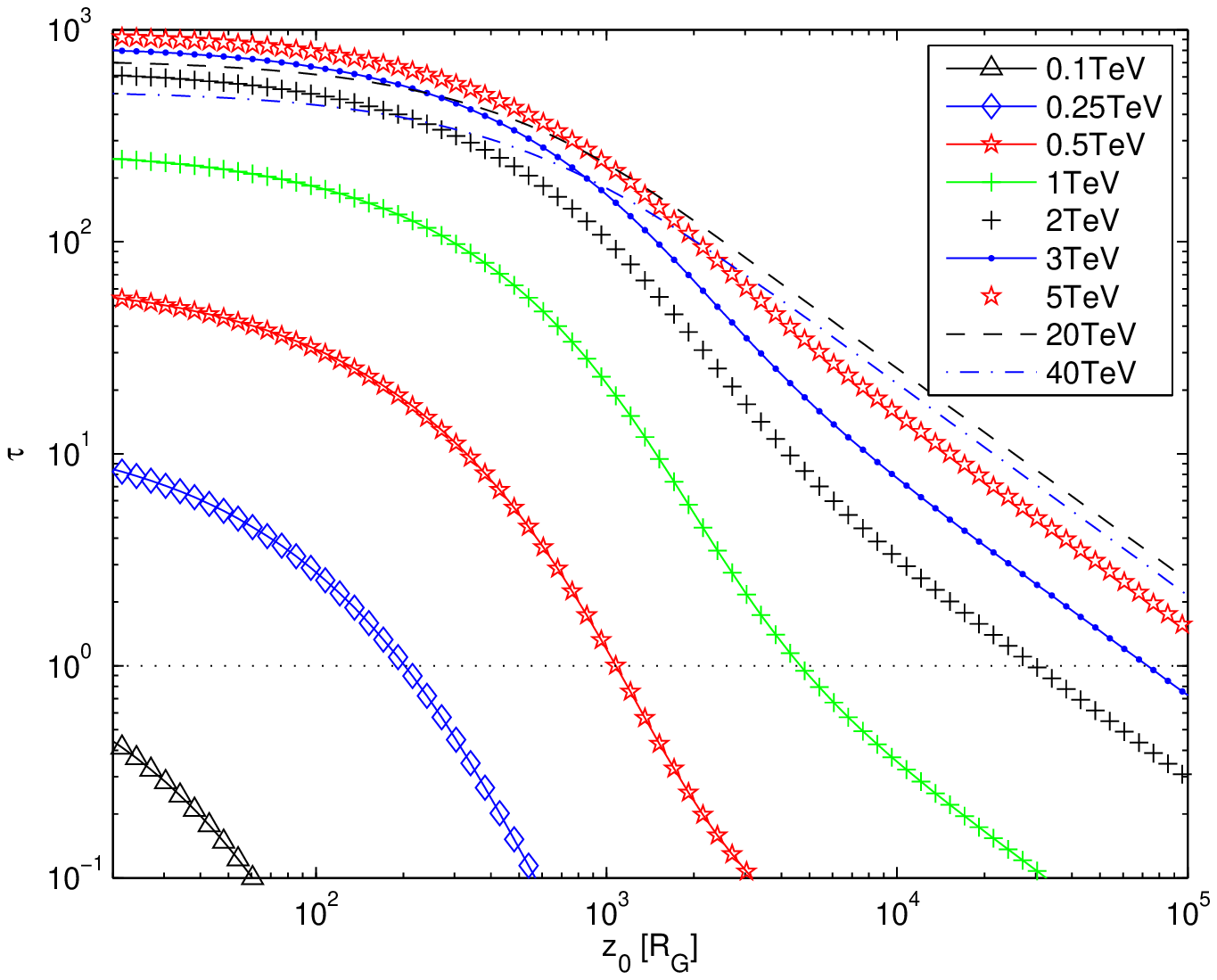} 
   \caption{\label{CenA_i80} \small Optical depth for TeV radiation produced in Cen\,A with inclination $i=\ang{80}$.}}
\end{figure}

\begin{figure}
\centering{
   \includegraphics[width=16.cm]{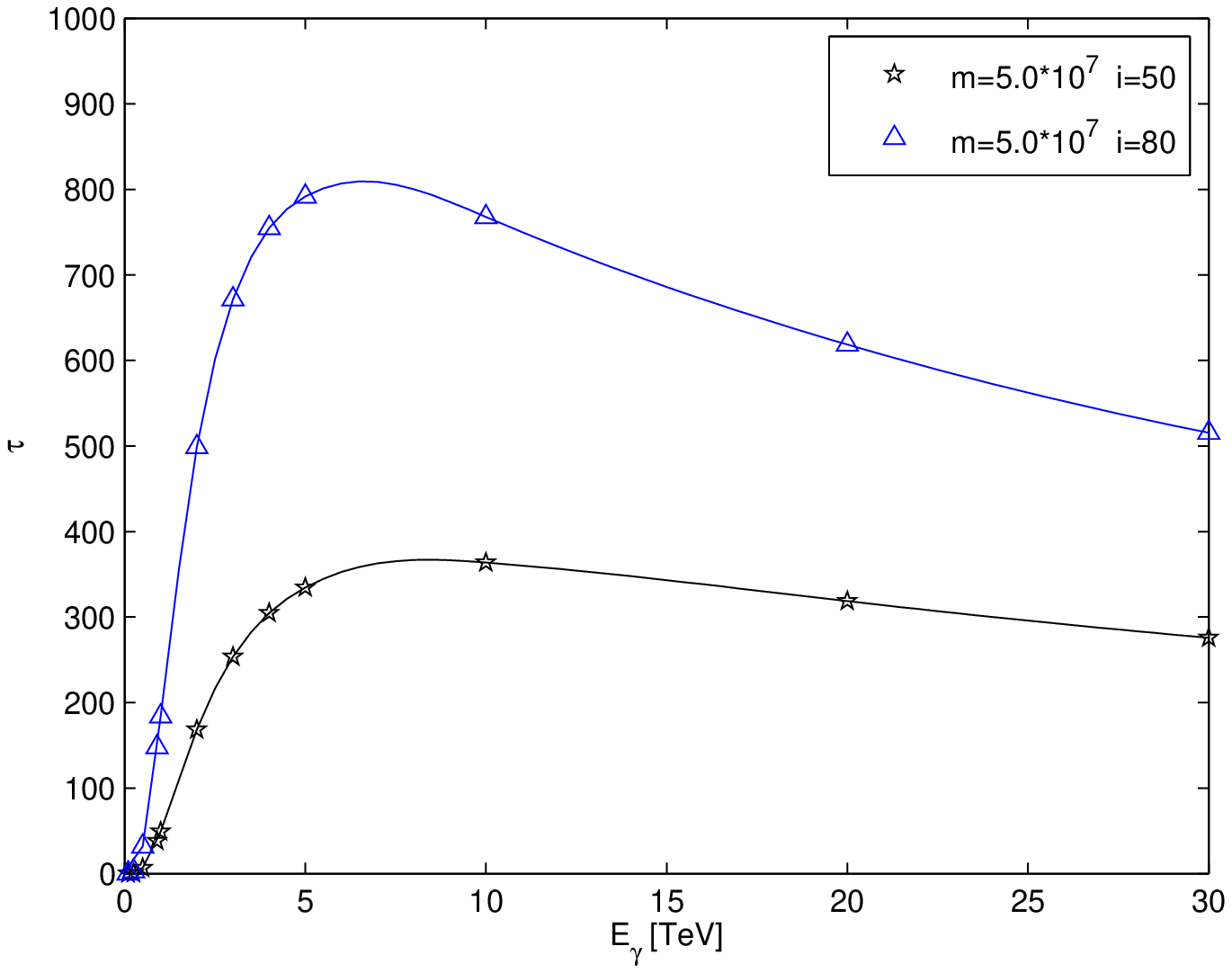}
   \caption{\label{CenA_100RG} \small Optical depth for TeV radiation produced in Cen\,A at $z_0=100\,R_G$.}}
\end{figure}

\begin{figure}
\centering{
   \includegraphics[width=16.cm]{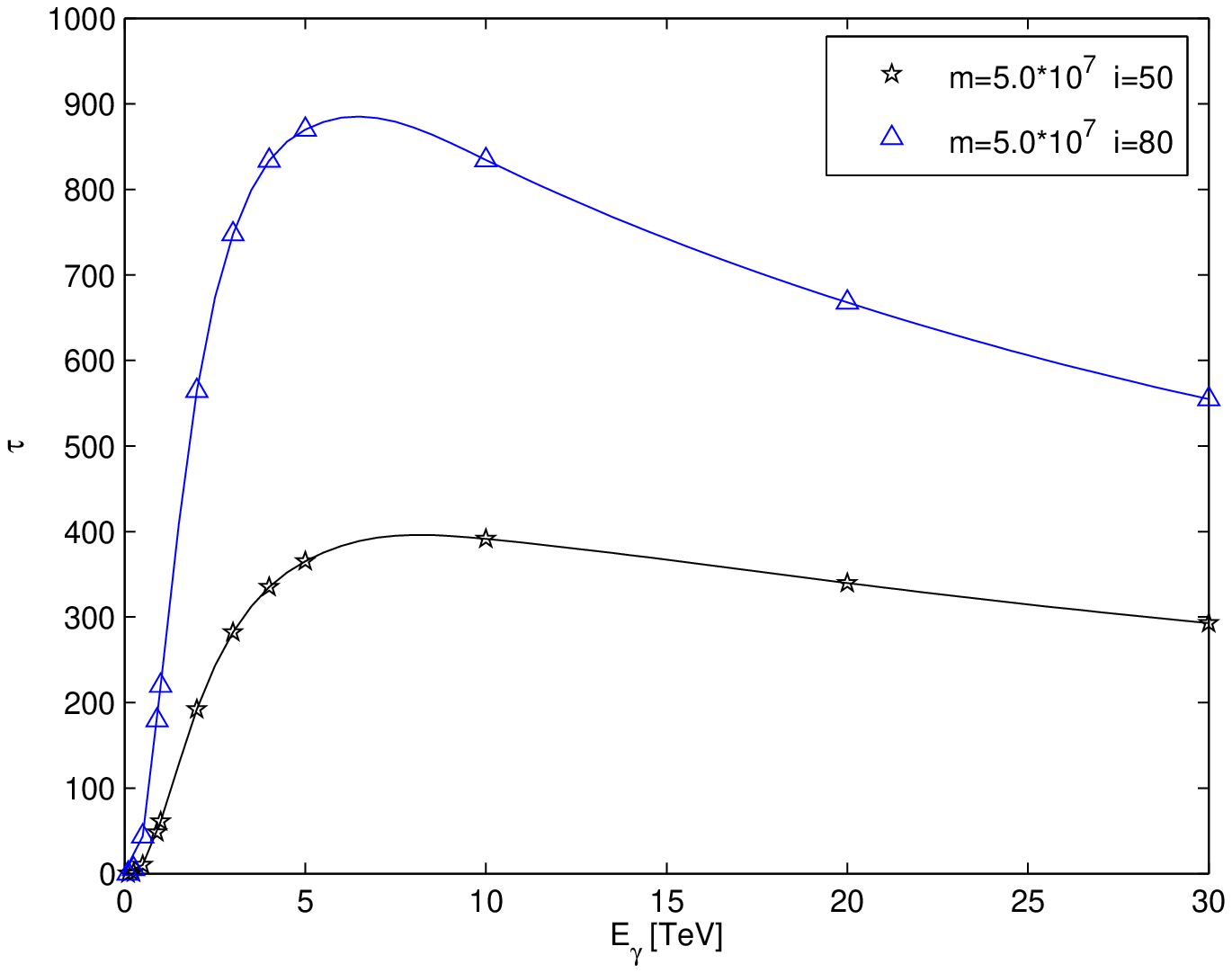} 
   \caption{\label{CenA_050RG} \small Optical depth for TeV radiation produced in Cen\,A at $z_0=50\,R_G$.}}
\end{figure}

Since we want to investigate if the observed $\gamma$ photons from Cen$\,$A can be produced in a similar way as the TeV photons from M87, we concentrate on the innermost $100\,R_G$ resp. $50\,R_G$, depending on the correct value of the M87 black hole mass. As can be seen from Figure \ref{CenA_i50}, the optical depth for an inclination of $i=\ang{50}$ and photon energies between $0.5$ and $40\,$TeV lies far above unity. An increase in inclination to $i=\ang{80}$ leads to an even stronger absorption of TeV photons (cf. Figure \ref{CenA_i80}), so $\gamma$ radiation from $0.25\,$TeV up to (at least) $40\,$TeV should be completely absorbed. Figures \ref{CenA_100RG} and \ref{CenA_050RG} give the optical depth as a function of the photon energy for both inclinations at a fixed production height of $100\,R_G$ and $50\,R_G$, respectively. It can be seen that the optical depth is much larger than one over a wide range of TeV energies. Moreover, the maximum absorption occurs for photons with $\approx 5\,$TeV (in case of $i= \ang{80}$) resp. $\approx 10\,$TeV (in case of $i= \ang{50}$). After this maximum the optical depth declines with increasing TeV photon energy but is still well above unity. Interestingly, this maximum value is more pronounced for $i=\ang{80}$ than for $i=\ang{50}$.

%xxxxxxxxxxxxxxxxxxxxxxxxxxxxxxxxxxxxxxxxxxxxxxxxxxxxxxxxxxxxxxxxxxxxxxxxxxxxxxxxxxxxxxxxxxxxxxxxxxxxxxxxxxxxxxxxxxxxx
%xxxxxxxxxxxxxxxxxxxxxxxxxxxxxxxxxxxxxxxxxxxxxxxxxxxxxxxxxxxxxxxxxxxxxxxxxxxxxxxxxxxxxxxxxxxxxxxxxxxxxxxxxxxxxxxxxxxxx
\section{Discussion} 
The results for Cen$\,$A are very different than for M87. Most notably, in Cen$\,$A the absorption of the TeV photons by disk photons is much higher than in M87. While the photon field produced by an SSD is optically thin for all observed TeV photon energies and nearly all considered parameters in M87, the SSD photons in Cen$\,$A lead to a total absorption of photons between $0.5$ and $40\,$TeV. Furthermore, for an emission distance of $100\,R_G$ resp. $50\,R_G$ the optical depth of Cen$\,$A attains its maximum at just $\approx 5\,$TeV (for $i=\ang{80}$) and $\approx 10\,$TeV (for $i=\ang{50}$), whereas in M87 the maximum absorption takes place at higher energies. The different results are traceable to the different values for the black hole mass, the bolometric disk luminosity and inclination of the object. %
Here, the influence of the term $L_{bol}/m^2$ is significant because it appears as an exponential function in the denominator of the optical depth (cf. eq.(\ref{opticaldepth})). Since its value is four orders of magnitude larger for Cen$\,$A than for M87, the optical depth in Cen$\,$A is orders of magnitude higher.
Moreover, due to a different black hole mass in M87 and Cen\,A, their SSD spectra are different. Because the highest temperature of an SSD lies near its inner radius, $r_{min}$, and the lowest temperature of the flow is achieved at the outer radius, $r_{max}$, radiation produced in the inner region is more energetic than radiation from the outer. In M87 the temperature of the material inside the SSD lies in the range of $\sim5 \cdot 10^4\,$K at $r=6$ and $\sim 500\,$K at $r=2\,000$. Since the maximum of the emission of a blackbody with the temperature $T$ lies at $E_{ph}=h\nu_{ph} = 2.82k_BT$, where $k_B$ is the Boltzmann constant, the inner edge of the M87 SSD emits mostly photons at $\sim (2-4)\,$eV, while the outer parts emit $\sim 0.05\,$eV photons. In Cen$\,$A the SSD spectrum is more energetic. The inner part of the disk has $T \sim 5 \cdot 10^5\,$K and therefore produces $\gamma$ radiation at $\sim 50\,$eV, while at large radii the temperature of the flow decreases to $\sim 10^4\,$K and the photon energy to $\sim1\,$eV. Resultantly, the $\gamma$ emission of the Cen\,A disk is primarily in the UV, while the SSD in M87 emits mostly in the optical and IR ranges. These differences lead to different absorption probabilities of TeV photons since photons between $1\,$eV and $50\,$eV have the highest interaction probabilities with $0.01\,$TeV$/(1-\mu)$ to $0.5\,$TeV$/(1-\mu)$ photons (Cen$\,$A). On the other hand, the SSD photons in M87 interact most readily with $0.1\,$TeV$/(1-\mu)$ to $10\,$TeV$/(1-\mu)$ photons. In these calculations the inclination of the galaxy in question is of great influence. Since M87 has an inclination below $\ang{45}$, small interaction angles ($\theta<\ang{90}$, $\mu=\cos{\theta}$) are more common than larger ones. This means that the maximum opacity for $\gamma$ photons occurs for increasingly high $\gamma$ energies, as evinced in our results -- we found that the TeV photon energy which maximum of the optical depth lies far above $10\,$TeV in M87. Cen$\,$A has a larger inclination than M87, favoring the occurrence of larger interaction angles ($\gtrsim \ang{90}$). Consequently, TeV photon absorption is highest for $E_{\gamma}=10\,$TeV ($i=\ang{50}$) resp. $E_{\gamma}=5\,$TeV ($i=\ang{80}$), \emph{ceteris paribus}. 

The results for the optical depth in M87 are consistent with observations: the SSD photon field is too faint to lead to a significant absorption of the observed TeV photons. Only if the black hole has a mass of $m=3.2 \cdot 10^9$ and the inclination is $\gtrsim \ang{30}$, $\gtrsim 20\,$TeV photons should undergo a notable attenuation. Even in this extreme case a portion of them should still be able to escape. Moreover, the assumed disk luminosity is only an upper limit and can be one (or more) order(s) of magnitude lower, leading to a smaller optical depth, which favours our model. Since the influence of a change in the disk luminosity can be illustrated by a change in the mass of the black hole (cf. Section \ref{Lbol}), we can simulate a decrease of the disk luminosity by an increase of the black hole mass. A change of $m=3.2 \cdot 10^9$ to $m=6.4 \cdot 10^9$ has a similar effect on the optical depth as a decrease of the disk luminosity to $75\%$ of the original assumed value. This variable replacement simulations produces results which are consistent with the estimation of the disk luminosity. On the whole, we can say that the absorption of TeV photons produced around $\sim100R_G$ resp. $\sim50R_G$ above the black hole is so small that the observed $\gamma$ radiation could come from the black hole vicinity.

On the other hand it is not possible to observe TeV $\gamma$ radiation produced within the innermost $\sim 100R_G$ of Cen$\,$A. Thus, either the location of the production site is different than in M87, or a low luminosity SSD does not explain the accretion in LLAGN appropriately.

%xxxxxxxxxxxxxxxxxxxxxxxxxxxxxxxxxxxxxxxxxxxxxxxxxxxxxxxxxxxxxxxxxxxxxxxxxxxxxxxxxxxxxx
% new paragraph (03.03.2011):

In this paper, we assume a relatively low accretion rate. In a model by \cite{BlandfordPayne} , it is proposed that AGN jets are disk-driven. This model was, for instance, used in \cite{BicknellWagner11} to estimate the optical depth for the blazar PKS 2155-304. Applying this model to M87 would predict a more luminous disk than assumed here, which would then be in contradiction to the detection of TeV emission within $100R_G$ of the central black hole of M87. Thus, the detection of TeV emission close to
the black hole of M87 challenges the disk-driven jet model and favors other models such as the Blandford-Znajek mechanism \citep{BlandfordZnajek} .

%xxxxxxxxxxxxxxxxxxxxxxxxxxxxxxxxxxxxxxxxxxxxxxxxxxxxxxxxxxxxxxxxxxxxxxxxxxxxxxxxxxxxxxxxxxxxxxxxxxxxxxxxxxxxxxxxxxxxx
%xxxxxxxxxxxxxxxxxxxxxxxxxxxxxxxxxxxxxxxxxxxxxxxxxxxxxxxxxxxxxxxxxxxxxxxxxxxxxxxxxxxxxxxxxxxxxxxxxxxxxxxxxxxxxxxxxxxxx
\section{Conclusions and Outlook}
We calculated the absorption probability for TeV photons produced in the vicinity of the supermassive black holes in the LLAGNs M87 and Cen$\,$A by photons produced by the accretion flows around each black hole. Our model assumed that the accretion flow can be described by a low luminosity Shakura-Sunyaev Disk, where the luminosity is set to the observed (low) luminosities in both objects. The calculations show that the results are very different between the two LLAGNs. While for M87 the SSD photon field is translucent for transversing TeV photons up to $\sim 15\,$TeV produced not more than $100R_G$ (for a black hole mass of $m=3.2 \cdot 10^9$) resp. $50R_G$ (for $m=6.4 \cdot 10^9$) away from the black hole, TeV photons produced near the black hole in Cen$\,$A are absorbed heavily. This implies that the core region in M87 is detectable at TeV energies whereas Cen$\,$A is not. The reason for this remarkable difference is the differing ratios of nuclear bolometric luminosity to the square of the black hole mass. This ratio is $5\cdot10^4 - 10^5$ times larger for Cen\,A than for M87, which is seen in the calculations to predict much higher absorption probabilities of TeV photons. Since both objects have been observed at TeV energies our calculations imply that either the production mechanism and production site of the TeV photons are different between the sources or that the assumption of a low luminosity SSD is not valid for accretion flows in LLAGNs. Our next step will be to carry out the calculations of the optical depth assuming different accretion models. An accretion disk model which does not depend so strongly on parameters like the black hole mass and the disk luminosity could produce similar results for both M87 and Cen$\,$A, thereby admitting TeV $\gamma$ production near the black holes in each galaxy. On the contrary, the black hole mass and disk luminosity are essential parameters for most accretion models, so it is questionable if they are in fact more insensitive to these two variables.
%xxxxxxxxxxxxxxxxxxxxxxxxxxxxxxxxxxxxxxxxxxxxxxxxxxxxxxxxxxxxxxxxxxxxxxxxxxxxxxxxxxxxxxxxxxxxxxxxxxxxxxxxxxxxxxxxxxxxx
%xxxxxxxxxxxxxxxxxxxxxxxxxxxxxxxxxxxxxxxxxxxxxxxxxxxxxxxxxxxxxxxxxxxxxxxxxxxxxxxxxxxxxxxxxxxxxxxxxxxxxxxxxxxxxxxxxxxxx

% figures:
%xxxxxxxxxxxxxxxxxxxxxxxxxxxxxxxxxxxxxxxxxxxxxxxxxxxxxxxxxxxxxxxxxxxxxxxxxxxxxxxxxxxxxxxxxxxxxxxxxxxxxxxxxxxxxxxxxxxxx
%____________________________________________________________________________________________________________
%____________________________________________________________________________________________________________

%____________________________________________________________________________________________________________

%____________________________________________________________________________________________________________

%____________________________________________________________________________________________________________

%____________________________________________________________________________________________________________

%____________________________________________________________________________________________________________

%____________________________________________________________________________________________________________
%____________________________________________________________________________________________________________
%____________________________________________________________________________________________________________
%____________________________________________________________________________________________________________
%____________________________________________________________________________________________________________
%____________________________________________________________________________________________________________
%xxxxxxxxxxxxxxxxxxxxxxxxxxxxxxxxxxxxxxxxxxxxxxxxxxxxxxxxxxxxxxxxxxxxxxxxxxxxxxxxxxxxxxxxxxxxxxxxxxxxxxxxxxxxxxxxxxxxx
%xxxxxxxxxxxxxxxxxxxxxxxxxxxxxxxxxxxxxxxxxxxxxxxxxxxxxxxxxxxxxxxxxxxxxxxxxxxxxxxxxxxxxxxxxxxxxxxxxxxxxxxxxxxxxxxxxxxxx
\acknowledgments
We wish to extend special thanks to Marek Abramowicz, \L ukasz Stawarz, Bo$\dot{\text{z}}$ena Czerny, Agnieszka Janiuk and Markus B\"ottcher for very useful suggestions and discussions. We also gratefully acknowledge support from the Research Department of Plasmas with Complex Interactions (Bochum) and from the DAAD (in particlar the RISE Programme), and partial support by the Deutsche Forschungsgemeinschaft (DFG Schl 201/20-1). Furthermore, we thank the anonymous referee for helpful comments.

\bibliography{biblPaper1}
\bibliographystyle{plainnat}
\end{document}